\documentclass[10pt,a4paper]{article}
\pdfoutput=1
\usepackage[english]{babel}
\usepackage[T1]{fontenc}
\usepackage[latin1]{inputenc}
\usepackage{amsfonts,amsbsy,bm,euscript,mathrsfs}
\usepackage{amssymb,stmaryrd,faktor,slashed,arydshln}
\usepackage[x11names]{xcolor}
\usepackage[tbtags]{amsmath}
\usepackage[bookmarks=true,colorlinks=true,linkcolor=black,citecolor=black,urlcolor=black,bookmarksnumbered]{hyperref}
\usepackage[nosort]{cite}
\usepackage{tikz,pgfplots}
\usetikzlibrary{matrix,fadings,calc,positioning,decorations.pathreplacing,arrows}
\usetikzlibrary{calc,trees,positioning,arrows,chains,shapes.geometric,decorations.pathreplacing,decorations.pathmorphing,shapes,matrix,shapes.symbols,shapes}

\numberwithin{equation}{section}

\makeatletter
\renewcommand\section{\@startsection {section}{1}{\z@}
{-3.5ex \@plus -1ex \@minus -.2ex}
{2.3ex \@plus.2ex}
{\normalfont\Large\bfseries}}
\renewcommand\subsection{\@startsection{subsection}{2}{\z@}
{-3.25ex\@plus -1ex \@minus -.2ex}
{1.5ex \@plus.2ex}
{\normalfont\large\bfseries}}
\makeatother

\newcommand{\arxivlink}[1]{\href{http://arxiv.org/abs/#1}{\tt arXiv:#1}}
\newcommand{\foot}[1]{\footnote{#1\vspace{2pt}}}

\newcommand{\Ad}{\operatorname{Ad}}
\newcommand{\Adi}{\operatorname{Ad}^{\vphantom{-1}}}
\newcommand{\AdS}{\operatorname{AdS}}
\newcommand{\diag}{\operatorname{diag}}
\newcommand{\Mat}{\operatorname{Mat}}
\newcommand{\GL}{\operatorname{GL}}
\newcommand{\SL}{\operatorname{SL}}
\newcommand{\PSL}{\operatorname{PSL}}
\newcommand{\SU}{\operatorname{SU}}
\newcommand{\SO}{\operatorname{SO}}
\newcommand{\U}{\operatorname{U}}
\newcommand{\PSU}{\operatorname{PSU}}
\newcommand{\Tr}{\operatorname{Tr}}
\newcommand{\tr}{\operatorname{tr}}
\newcommand{\p}{\text{p}}
\newcommand{\e}{\text{e}}
\newcommand{\scale}{\text{R}}

\newcommand{\sdp}{\mu}
\newcommand{\sdpc}{\kappa}
\newcommand{\tl}{\tilde \lambda}

\DeclareMathOperator{\arcsinh}{arcsinh}

\newenvironment{psmallmatrix}{\left(\begin{smallmatrix}}{\end{smallmatrix}\right)}

\begin{document}

\setcounter{equation}{0}
\setcounter{footnote}{0}
\setcounter{section}{0}

\thispagestyle{empty}

\begin{flushright} \texttt{HU-EP-16/14\\HU-MATH-16/09}\end{flushright}

\begin{center}
\vspace{1.5truecm}

{\LARGE \bf Non-split and split deformations of AdS$_{\mathbf{5}}$}

\vspace{1.5truecm}

{Ben Hoare$^{1}$ and Stijn J. van Tongeren$^{2}$}

\vspace{1.0truecm}

{\em $^{1}$ Institut f\"ur Theoretische Physik, ETH Z\"urich,\\ Wolfgang-Pauli-Strasse 27, 8093 Z\"urich, Switzerland}

\vspace{0.5truecm}

{\em $^{2}$ Institut f\"ur Mathematik und Institut f\"ur Physik, Humboldt-Universit\"at zu Berlin, \\ IRIS Geb\"aude, Zum Grossen Windkanal 6, 12489 Berlin, Germany}

\vspace{1.0truecm}

{{\tt bhoare@ethz.ch, \quad svantongeren@physik.hu-berlin.de}}

\vspace{1.0truecm}
\end{center}

\begin{abstract}
The $\eta$ deformation of the $\AdS_5 \times S^5$ superstring depends on a non-split $r$ matrix for the superalgebra $\mathfrak{psu}(2,2|4)$. Much of the investigation into this model has considered one particular choice, however there are a number of inequivalent alternatives. This is also true for the bosonic sector of the theory with $\mathfrak{su}(2,2)$, the isometry algebra of $\AdS_5$, admitting one split and three non-split $r$ matrices. In this article we explore these $r$ matrices and the corresponding geometries. We investigate their contraction limits, comment on supergravity backgrounds and demonstrate their relation to gauged-WZW deformations. We then extend the three non-split cases to $\AdS_5 \times S^5$ and compute four separate bosonic two-particle tree-level $S$-matrices based on inequivalent BMN-type light-cone gauges. The resulting $S$-matrices, while different, are related by momentum-dependent one-particle changes of basis.
\end{abstract}

\newpage

\setcounter{equation}{0}
\setcounter{footnote}{0}
\setcounter{section}{0}

\tableofcontents

\section{Introduction}\label{sec:intro}

The $\eta$ deformation of the $\AdS_5 \times S^5$ superstring \cite{Delduc:2013qra,Delduc:2014kha} is a deformation of the semi-symmetric space supercoset action of \cite{Metsaev:1998it,Berkovits:1999zq} with $q$-deformed $\mathfrak{psu}(2,2|4)$ symmetry \cite{Delduc:2014kha}. The construction generalises the deformation of two bosonic models, the principal chiral \cite{Klimcik:2002zj,Klimcik:2008eq} and the symmetric space sigma model \cite{Delduc:2013fga}. The theory remains integrable and has a form of $\kappa$ symmetry. As such it was a natural candidate to describe a type IIB Green-Schwarz superstring, recovering the familiar $\AdS_5 \times S^5$ model in the undeformed limit.

Whether this deformation of the $\AdS_5 \times S^5$ superstring is itself a superstring theory has by now been thoroughly investigated. The first approach taken was to determine the metric and $B$-field \cite{Arutyunov:2013ega} and, investigating certain limits and truncations \cite{Hoare:2014pna,Arutynov:2014ota,Arutyunov:2014cra}, attempt to find a dilaton and R-R fluxes that complete the NS-NS fields to a type IIB supergravity solution \cite{Lunin:2014tsa}. This approach was partially successful, however it was unclear whether the resulting backgrounds matched the expansion of the supercoset action or corresponded to an integrable worldsheet sigma model. In \cite{Arutyunov:2015qva} the deformed semi-symmetric space supercoset action was expanded to quadratic order in fermions and the R-R fluxes extracted. It transpired that there is no dilaton that completes the NS-NS fields and R-R fluxes to a type IIB supergravity solution.

While this story has been partially understood by considering the T-dual background, which is a type IIB$^*$ supergravity solution, albeit with a dilaton breaking the isometries \cite{Hoare:2015wia,Arutyunov:2015mqj}, there are other puzzling aspects to the results of \cite{Arutyunov:2015qva}.

The first is the so-called mirror, or contraction, limit, in which the deformation parameter is taken to infinity. Taking this limit in the deformed metric and $B$-field gives the T-dual of $\textrm{dS}_5 \times H^5$ with vanishing $B$-field \cite{Arutyunov:2014cra}. This background matches the bosonic background of the mirror $\AdS_5 \times S^5$ superstring, the background of the sigma model that is related to the $\AdS_5 \times S^5$ one by a worldsheet double Wick rotation once a light cone gauge is fixed. This limit was inspired by the mirror duality of the exact $q$-deformed $S$-matrix \cite{Arutynov:2014ota} -- meaning its invariance under inversion of the deformation parameter combined with a mirror transformation (double Wick rotation) -- which at a geometric level extends to the deformed bosonic background \cite{Arutyunov:2014cra}. Based on this structure a conjecture was made for the form of the R-R fluxes and dilaton in this maximal deformation limit \cite{Arutyunov:2014cra}, matching the double Wick rotation of the $\AdS_5 \times S^5$ fermions \cite{Arutyunov:2014jfa}. However, the limit of the R-R fluxes found in \cite{Arutyunov:2015qva} does not match the conjecture of \cite{Arutyunov:2014cra} and furthermore does not respect the geometric mirror duality of \cite{Arutyunov:2014cra,Arutyunov:2014jfa} which would naturally imply the algebraic one of the exact $S$-matrix \cite{Arutynov:2014ota}. The converse -- mirror duality of the $S$-matrix implying geometric mirror duality -- does not hold, but it is natural to expect it to be the case. This is related to the second puzzle just below. Algebraically, this maximal deformation limit can be understood as a contraction of the $q$-deformed symmetry algebra to a $\kappa$-Poincar\'e type algebra \cite{Pachol:2015mfa}. In particular, in this limit the symmetry algebra of the $\eta$-deformed $\AdS_5 \times S^5$ superstring is a contraction of the $q$-deformed $\mathfrak{psu}(2,2|4)$ that contains the light-cone gauge symmetry of the mirror model as a subalgebra \cite{Pachol:2015mfa}.

The second puzzle relates to the light-cone gauge $S$-matrix itself.
Considering a BMN-type light-cone gauge, the two-particle tree-level $S$-matrix elements for two bosons and two fermions were computed in \cite{Arutyunov:2015qva}. The four boson amplitudes were computed earlier in \cite{Arutyunov:2013ega}. As a consequence of the new amplitudes the tree-level $S$-matrix no longer satisfies the classical Yang-Baxter equation. A two-particle change of basis was constructed that relates the tree-level result to the $S$-matrix that follows from symmetries and satisfies the classical Yang-Baxter equation. However, it remains to be understood if this change of basis can be extended to arbitrary orders and an arbitrary number of excitations.

\

The deformed model depends on a constant antisymmetric solution of the modified classical Yang-Baxter equation (mcYBe) for the superalgebra $\mathfrak{psu}(2,2|4)$. A particular solution was considered in \cite{Arutyunov:2013ega} and henceforth much of the investigation into the deformation has been based on this choice. However, there are other inequivalent options. The aim of this paper is to study the effect of considering these different choices on both the background geometries and the light-cone gauge $S$-matrices, in particular to see if they could help to resolve the puzzles of \cite{Arutyunov:2015qva}

We will focus on the deformed bosonic $\AdS_5 \times S^5$ model.
Therefore the action of interest is the deformation of the symmetric space sigma model for the coset $\frac{G}{H}$ \cite{Delduc:2013fga}
\begin{equation}\begin{split}\label{def:non-split}
\mathcal{S} = \frac{T}{2} \int d\tau d\sigma \, (\sqrt{h} h^{\alpha \beta} -\epsilon^{\alpha \beta}) \Tr (A_\alpha P \frac{1}{1 - \varkappa R_g P} A_\beta) \ , \qquad A & = g^{-1} dg  \ , \qquad  R_g = \Ad_g^{-1} R \Adi_g \ .
\end{split}\end{equation}
Here $T$ is the would-be effective string tension, $h$ is the worldsheet metric with $\sqrt{h}h^{\tau\tau} < 0$ and $\epsilon^{\tau\sigma}=1$. The field $g$ takes values in the group $G$ and $\Tr$ is an appropriately normalised invariant bilinear form. As $\frac{G}{H}$ is a symmetric space, $\mathfrak{g} = Lie(G)$ has a $\mathbb{Z}_2$ outer automorphism, with the grade $0$ space given by the subalgebra $\mathfrak{h} = Lie(H)$. $P$ then denotes the projector onto the grade $1$ space.

The operator $R$ is a linear map from $\mathfrak{g}$ to itself.
Provided it is antisymmetric,
\begin{equation}
\mathrm{Tr}(R(m) n) = -\mathrm{Tr}(m R(n)) \ ,
\end{equation}
and satisfies the non-split or split mcYBe
\begin{align}
\label{eq:cYBe}
& [R(m),R(n)] - R([R(m),n] + [m,R(n)])=\pm[m,n] \ ,
\end{align}
the deformed model is classically integrable. Here the plus sign corresponds to non-split and the minus sign to split. In the non-split case we use the deformation parameter $\varkappa$ of \cite{Arutyunov:2013ega}, which is related to the parameter $\eta$ of \cite{Delduc:2013fga} by $\varkappa = \eta$ and to that of \cite{Delduc:2013qra} by $\varkappa = 2\eta/ (1-\eta^2)$. In the split case we use the deformation parameter $\sdp$. Without loss of generality we will assume these to be positive. When these parameters are set to zero we recover the action for the symmetric space sigma model as expected.

Using the bilinear form the operator $R$ can be represented by an $r$ matrix, i.e.
\begin{equation}\label{Rrmap}
R(m) = \mathrm{Tr}_2(r (1\otimes m)) \ ,
\end{equation}
with
\begin{equation}
r = \sum \alpha_{ij} t^i \wedge t^j \equiv \sum \alpha_{ij} (t^i \otimes t^j - t^j \otimes t^i) \in \mathfrak{g} \otimes \mathfrak{g} \ ,
\end{equation}
where the $t^i$ are generators of $\mathfrak{g}$ and the $\alpha_{ij} \in \mathbb{R}$. In this article we will refer to both the operator $R$ and its matrix representation $r$ as the $r$ matrix, where the latter satisfies the mcYBe in the form
\begin{equation}\label{rrmcybe}
[r_{12},r_{13}]+[r_{12},r_{23}]+[r_{13},r_{23}] = \pm \Omega \ .
\end{equation}
Here $r_{mn}$ denotes the matrix realisation of $r$ acting in spaces $m$ and $n$ of the triple tensor product, while $\Omega$ is the canonical invariant element of $\Lambda^3_\mathfrak{g}$ with its overall scale fixed such that the plus (respectively minus) sign in \eqref{rrmcybe} corresponds to the non-split (respectively split) mcYBe.

\

The $r$ matrices we study in this article are those for $\mathfrak{su}(2,2)$ and $\mathfrak{su}(4)$, the isometry algebras of $\AdS_5$ and $S^5$ respectively. Furthermore, we will focus on inequivalent solutions of the mcYBe, that is up to inner automorphisms, and also up to compatible solutions of the classical Yang-Baxter equation (cYBe)
\begin{equation}
[R(m),R(n)] - R([R(m),n] + [m,R(n)])= 0 \ .
\end{equation}
In particular, we fix the latter requirement by the condition $R^3 = \mp R$ (where the minus sign corresponds to the non-split case and the plus sign to the split case).

Both $\mathfrak{su}(2,2)$ and $\mathfrak{su}(4)$ are real forms of the complex Lie algebra $\mathfrak{sl}(4;\mathbb{C})$. To the best of our knowledge \cite{Delduc:2014kha,CahenGuttRawnsley} there are three inequivalent solutions of the non-split and one of the split mcYBe for the real form $\mathfrak{su}(2,2)$. In contrast, for the compact real form $\mathfrak{su}(4)$ there is a single solution of the non-split mcYBe and no split solutions. Therefore, while the three non-split $\mathfrak{su}(2,2)$ $r$ matrices can be extended to inequivalent solutions for $\mathfrak{psu}(2,2|4)$, the same is not true for the split case.
Even so, considering the latter is still useful due to the existence of non-trivial limits in which it solves the cYBe. Such limits also exist for the non-split $r$ matrices. These $r$ matrices can be extended to solutions of the cYBe for $\mathfrak{psu}(2,2|4)$, which can also be used to define deformations of the $\AdS_5 \times S^5$ superstring \cite{Kawaguchi:2014qwa,Matsumoto:2014nra,Matsumoto:2015jja,vanTongeren:2015soa,Vicedo:2015pna,vanTongeren:2015uha}. The limits turn out to be particularly useful for investigating in which cases these deformations correspond to type IIB superstrings \cite{Hoare:2016hwh}.

This class of deformations of the principal chiral, symmetric space and semi-symmetric space sigma models based on solutions to the cYBe and mcYBe are referred to as Yang-Baxter deformations \cite{Vicedo:2015pna}. Here, to distinguish these two cases we will refer to those based on the mcYBe (respectively cYBe) as inhomogeneous (respectively homogeneous) Yang-Baxter deformations. We will also refer to the specific model studied in \cite{Arutyunov:2013ega} and subsequent papers as the $\eta$ deformation.

We consider the three non-split solutions and one split solution of the mcYBe for $\mathfrak{su}(2,2)$ and the corresponding deformations of $\AdS_5$. In section \ref{sec:deformations}, for each $r$ matrix we extract the geometry and $B$-field, with the sign of the latter defined by
\begin{equation}
\mathcal{S} = -\tfrac{T}{2}\int d\tau d\sigma \, (\sqrt{h}h^{\alpha\beta}G_{MN} - \epsilon^{\alpha\beta}B_{MN})\partial_\alpha x^M \partial_\beta x^N \ .
\end{equation}
Furthermore, in the non-split case we show that the three backgrounds are related by real diffeomorphisms, although there are still three regions separated by singularities. The section is concluded with three more general discussions. In the first we explore the contraction limits of the deformed models. The second outlines how one could use analytic continuations and the results of \cite{Arutyunov:2015qva,Hoare:2015wia} to conjecture the corresponding R-R fluxes for the various $r$ matrices, while the third demonstrates, for the $\AdS_3$ case, how the different backgrounds appear as limits of the gauged-WZW deformation \cite{Sfetsos:2013wia,Hollowood:2014rla,Hoare:2015gda}. In section \ref{sec:smat} we return to $\AdS_5 \times S^5$ and the three non-split $r$ matrices. Computing the bosonic two-particle tree-level $S$-matrices for four inequivalent BMN-type light-cone gauges (discussed in the appendix) we find differing results. However the $S$-matrices are related by momentum-dependent one-particle changes of basis. We finish in section \ref{sec:conclusions} with some concluding remarks and open questions.

\section{Non-split and split deformations of \texorpdfstring{$\mathbf{AdS_5}$}{AdS5}}\label{sec:deformations}

In this section we will discuss the various inhomogeneous Yang-Baxter deformations, non-split and split, of $\AdS_5$. The isometry group of $\AdS_5$ is $\SO(2,4)$. In each case the deformation only preserves a Cartan subgroup, i.e. three commuting isometries. In the non-split case this symmetry is $\U(1)^3$ while in the split case it is $\mathbb{R}^2 \times \U(1)$.

The defining matrix representation of the algebra $\mathfrak{su}(2,2) \simeq \mathfrak{so}(2,4)$ can be taken as
\begin{equation}\label{su22rep}
m \in \Mat(4;\mathbb{C}) \ , \qquad \Tr(m) = 0 \ , \qquad m^\dagger = m^* \equiv - \Adi_{\gamma^5} m \ , \qquad \gamma^5 = \diag(1,1,-1,-1) \ .
\end{equation}
for which we use the following basis \cite{Arutyunov:2009ga,Pachol:2015mfa}
\begin{equation}\label{mgen}
m^{ij} = \frac14 [\gamma^i,\gamma^j] \ , \qquad   m^{i5} = - m^{5i} = \frac12 \gamma^i \ , \qquad i = 0,\ldots,4 \ ,
\end{equation}
where
\begin{align}
\nonumber
\gamma^0 & = i \sigma_3 \otimes \sigma_0 \ , & \gamma^1 & = \sigma_2 \otimes \sigma_2 \ , & \gamma^2 & = - \sigma_2 \otimes \sigma_1 \ ,
\\
\gamma^3 & = \sigma_1 \otimes \sigma_0 \ , & \gamma^4 & = \sigma_2 \otimes \sigma_3 \ , & \gamma^5 & = - i \gamma^0 \ ,
\end{align}
$\sigma_0 = \mathbf{1}_{2\times 2}$ and $\sigma_a$ are the Pauli matrices.
The generators $m^{ij}$ then satisfy the standard $\mathfrak{so}(2,4)$ commutation relations
\begin{equation}
[m^{ij},m^{kl}] = \eta^{jk}m^{il} - \eta^{ik}m^{jl} - \eta^{jl} m^{ik} + \eta^{il}m^{jk} \ , \qquad i,j,k,l = 0,\ldots, 5 \ ,
\end{equation}
where $\eta = \operatorname{diag}(-1,1,1,1,1,-1)$ is the metric with which we raise and lower indices.

The symmetric space $\AdS_5$ is equivalent to the coset $\frac{\SO(2,4)}{\SO(1,4)}$. We take the gauge algebra $\mathfrak{so}(1,4)$ to be generated by $m^{ij}$, $i,j=0,\ldots,4$, and hence $P$ in \eqref{def:non-split} projects onto the space spanned by $m^{i5}$.

\subsection{Non-split deformations}\label{ssec:non-split}

\subsubsection*{r matrices}

Our starting point for the discussion of non-split $r$ matrices for $\mathfrak{su}(2,2)$, equivalently $\mathfrak{so}(2,4)$, and the associated deformations of $\AdS_5$ is the case much studied in the literature \cite{Delduc:2013qra,Delduc:2014kha,Delduc:2013fga,Arutyunov:2013ega,Hoare:2014pna,Arutynov:2014ota,Arutyunov:2014cra,Hoare:2015gda,Arutyunov:2015qva,Hoare:2015wia,Arutyunov:2015mqj}.

Decomposing a matrix $m$ taking values in the $\mathfrak{su}(2,2)$ representation \eqref{su22rep} into its diagonal and upper and lower triangular parts (denoted $m_d$, $m_+$ and $m_-$ respectively), the standard $r$ matrix acts as
\begin{equation}\label{canonical}
R_0(m_d + m_+ + m_-) = - i m_+ + i m_- \ ,
\end{equation}
which one can check preserves the defining relations \eqref{su22rep}, i.e. $\Tr(R_0(m)) = 0$ and $R_0(m)^\dagger = R_0(m)^*$ for $m \in \mathfrak{su}(2,2)$.

New $r$ matrices can then be constructed from the canonical one using a constant element $X \in \SL(4;\mathbb{C})$\foot{In principle one could consider $X \in \GL(4;\mathbb{C})$, however elements proportional to the identity will give the same $r$ matrix.} (given here by invertible elements of $\Mat(4;\mathbb{C})$ with unit determinant)
\begin{equation}\label{rxrxns}
R_{X} = \Ad_{X}^{-1} R \Adi_{X} \ , \qquad r_X = (\Ad_X^{-1} \otimes \Ad_X^{-1}) \, r \ ,
\end{equation}
For the resulting $R_X$ to be a non-split $r$ matrix for $\mathfrak{su}(2,2)$, it should preserve the defining relations \eqref{su22rep}. Furthermore, we say two $r$ matrices are equivalent if $X \in \SU(2,2)$ since the corresponding transformation can then be understood as an inner automorphism of the algebra. Alternatively in the deformed action \eqref{def:non-split} the matrix $X$ can be absorbed into the definition of the group-valued field $g$. The problem of classifying inequivalent non-split $r$ matrices for non-compact real forms is an open question. Therefore, here we will consider the three inequivalent non-split $r$ matrices for the algebra $\mathfrak{su}(2,2)$ discussed in \cite{Delduc:2014kha}. These are given by $R_{1,2} \equiv (R_0)_{P_{1,2}}$ where
\begin{equation}\label{p1p2}
P_1 = \begin{psmallmatrix} 1 & 0 & 0 & 0 \\ 0 & 0 & i & 0 \\ 0 & i & 0 & 0 \\ 0 & 0 & 0 & 1 \end{psmallmatrix} = \exp\big[-\frac{i\pi}2 (m^{02} - m^{15})\big] \ , \qquad
P_2 = \begin{psmallmatrix} 1 & 0 & 0 & 0 \\ 0 & 0 & 0 & i \\ 0 & 0 & 1 & 0 \\ 0 & i & 0 & 0 \end{psmallmatrix} = \exp\big[-\frac{i\pi}2 (m^{04} - m^{35})\big] \ , \qquad
\end{equation}
From these expressions we can immediately see that $P_1$ and $P_2$ are not in $\SU(2,2)$, however, one can check they do preserve the defining relations \eqref{su22rep}. They are related to the permutations of \cite{Delduc:2014kha} by $\SU(2,2)$ transformations and multiples of the identity. One can check that their action on the signature matrix $\gamma_5$ takes the expected form
\begin{equation}
\Ad_{P_1}^{-1}\gamma_5 = \diag(1,-1,1,-1) \ , \qquad
\Ad_{P_2}^{-1}\gamma_5 = \diag(1,-1,-1,1) \ .
\end{equation}
Written in terms of the basis \eqref{mgen}, the $r$ matrices associated to the operators $R_0$, $R_1$ and $R_2$ are
\begin{align}
\nonumber
r_{0} & = -(m^{0}{}_{i}\wedge m^{i5}+ m^{1}{}_{3}\wedge m^{32}+m^{1}{}_{4}\wedge m^{42}) \ , \\
\nonumber
r_{1} & = -(m^{1}{}_{i} \wedge m^{i2} + m^{0}{}_{3} \wedge m^{35} + m^{0}{}_{4} \wedge m^{45}) \ , \\
r_{2} & = -(m^{3}{}_{i} \wedge m^{i4} + m^{1}{}_{0} \wedge m^{02} + m^{1}{}_{5} \wedge m^{52}) \ .
\label{eq:non-splitrmatrices}
\end{align}

Another way to arrive at these $r$ matrices is to start from the unique $\mathfrak{sl}(4;\mathbb{C})$ $r$ matrix of the form $e \wedge f$, where $e$ and $f$ are positive and negative roots, and fix the real form $\mathfrak{su}(2,2)$, which can be done in multiple inequivalent ways. In a twist on this picture, we can first fix the compact real form $\mathfrak{su}(4)$ -- for which there is only one non-split $r$ matrix up to inner automorphisms -- and analytically continue to $\mathfrak{su}(2,2)$.\footnote{This idea was explored for $\mathfrak{sl}(2;\mathbb{C})$ in \cite{Pachol:2015mfa}.} In our conventions this $r$ matrix is
\begin{equation}
\label{eq:su4rmat}
r_{\mathfrak{su}(4)} =-(n^{5}{}_{i} \wedge n^{i6} + n^{1}{}_{3} \wedge n^{32} +n^{1}{}_{4} \wedge n^{42}) \ ,
\end{equation}
where $n^{ij}$, $i,j = 1,\ldots, 6$, are the generators of $\mathfrak{su}(4)$. It arises from the $\mathfrak{sl}(4;\mathbb{C})$ $r$ matrix by identifying the roots of $\mathfrak{sl}(4;\mathbb{C})$ in terms of the generators $n^{ij}$. This can be done in many ways as any permutation of the indices $1,\ldots,6$ or $\mathrm{SU}(4)$ transformation gives an admissible $r$ matrix. However, the former leave the $r$ matrix \eqref{eq:su4rmat} invariant up to inner automorphisms, while the latter are just inner automorphisms themselves. Therefore, in contrast to the non-compact case these choices do not affect the $r$ matrix.

To analytically continue from $\mathfrak{su}(4) \simeq \mathfrak{so}(6)$ to $\mathfrak{su}(2,2) \simeq \mathfrak{so}(2,4)$ we can think in terms of the $\mathbb{R}^6$ associated to $\mathfrak{so}(6)$, and analytically continue a two-plane to obtain the $\mathbb{R}^{2,4}$ of $\mathfrak{so}(2,4)$. Note that up to inner automorphisms (and hence permutations of indices) of the $\mathfrak{su}(4)$ $r$ matrix we can choose this two-plane to be associated to the span of any two indices, where analytic continuation corresponds to ``multiplying these indices by $i$''. We need this continuation to result in a real $r$ matrix, so that we obtain a real non-split $r$ matrix for $\mathfrak{su}(2,2)$. Starting from \eqref{eq:su4rmat} this can be done in three inequivalent ways: if we analytically continue $5 \rightarrow i 0$ we are forced to choose $6 \rightarrow i 5$, if we continue $1 \rightarrow i 0$ we are forced to choose $2 \rightarrow i 5$, and finally if we continue $3 \rightarrow i 0$ we also need $4 \rightarrow i 5$. Up to inconsequential signs and reshuffling of indices within the timelike and spacelike sets, this gives precisely the three $r$ matrices \eqref{eq:non-splitrmatrices} and nothing more. While it seems unlikely there are further inequivalent non-split $r$ matrices, we have not proven that this procedure is exhaustive.

\subsubsection*{Geometries}

For the $r$ matrix $r_0$ we use the parametrisation of the $\frac{SO(2,4)}{SO(1,4)}$ coset of \cite{Arutyunov:2013ega}
\begin{equation}
\label{eq:gns0}
g_0 = \lambda e^{ \arcsin{x} \, m^{13}} e^{ \arcsinh{\rho}\, m^{15}} \ ,\\
\end{equation}
where
\begin{equation}\label{lambdaofns}
\lambda =  e^{t m^{05} - \psi_1 m^{12} - \psi_2 m^{34}} \ ,
\end{equation}
with isometric coordinates $t$, $\psi_1$ and $\psi_2$, and non-isometric coordinates $\rho$ and $x$
\begin{equation}\label{ranges0}
\rho \in [0,\infty) \ , \qquad x \in [0,1) \ .
\end{equation}
As observed in \cite{Delduc:2014kha}, for the $r$ matrices $r_1$ and $r_2$ this choice of parametrisation leads to a non-diagonal metric. Here we instead use the coset parametrisations\foot{The alternate choice of parametrisation of $g_1$ is obtained from the original \eqref{eq:gns0} by permuting indices $0$ with $2$ and $5$ with $1$, matching the relation between the $r$ matrices \eqref{p1p2}. Regarding $g_2$ we first note that the corresponding permutation, i.e. exchanging $0$ with $4$ and $5$ with $3$ in $g_0$, does not provide a good representation of the coset. However, using the gauge and global symmetry we can find an alternative form of $g_0$
\begin{equation*}
\lambda e^{ \arccos{x} \, m^{13}} e^{ \arcsinh{\rho}\, m^{35}} \ ,
\end{equation*}
such that the permutation does give a good representation.}
\begin{align}
\nonumber
g_1= & \lambda e^{\arcsinh{x} \, m^{35}} e^{\arcsinh{\rho} \, m^{15}} \ , \\
\label{eq:gns12}
g_2= & \lambda e^{\arcsinh{x} \, m^{15}} e^{\arcsinh{\rho} \, m^{35}} \ ,
\end{align}
which directly lead to diagonal metrics. Here we have labelled the coordinates as before, however the non-isometric coordinates are now both non-compact
\begin{equation}\label{ranges12}
\rho \in [0,\infty) \ , \qquad x \in [0,\infty) \ .
\end{equation}

For the $r$ matrix $r_0$ the resulting metric and $B$-field are given by
\begin{align}
\nonumber
ds_0^2 & = \frac1{1-\varkappa^2\rho^2}\Big( - (1+\rho^2) dt^2 + \frac{d\rho^2}{1+\rho^2} \Big)
          + \frac1{1+\varkappa^2\rho^4x^2}\Big( \rho^2 (1-x^2) d\psi_1^2 + \frac{\rho^2 dx^2}{1-x^2}\Big) + \rho^2 x^2 d\psi_2^2
\\ \label{backk0}
B_0 & = \frac{\varkappa\rho}{1-\varkappa^2\rho^2} dt \wedge d\rho + \frac{\varkappa \rho^4 x}{1+\varkappa^2 \rho^4 x^2} d\psi_1 \wedge dx \ ,
\end{align}
while for $r_1$ they take the form
\begin{align}
\nonumber
ds_1^2 & =  \frac1{1+\varkappa^2(1+\rho^2)}\Big(\rho^2 d\psi_1^2 + \frac{d\rho^2}{1+\rho^2}\Big)
\\ \nonumber & \quad + \frac1{1-\varkappa^2(1+\rho^2)^2x^2}\Big(-(1+\rho^2)(1+x^2)dt^2 + \frac{(1+\rho^2)dx^2}{1+x^2}\Big) + (1+\rho^2)x^2 d\psi_2^2
\\ \label{backk1}
B_1 & = \frac{\varkappa \rho}{1+\varkappa^2(1+\rho^2)} d\psi_1 \wedge d\rho + \frac{\varkappa (1+\rho^2)^2 x}{1-\varkappa^2(1+\rho^2)^2 x^2} dt \wedge dx \ ,
\end{align}
and finally for $r_2$ we find
\begin{align}
\nonumber
ds_2^2 & = \frac1{1+\varkappa^2(1+\rho^2)}\Big(\rho^2 d\psi_2^2 + \frac{d\rho^2}{1+\rho^2}\Big)
\\ \nonumber & \quad +
 \frac1{1+\varkappa^2 (1+\rho^2)^2(1+x^2)}\Big((1+\rho^2)x^2 d\psi_1^2 + \frac{(1+\rho^2)dx^2}{1+x^2} \Big)
-(1+\rho^2)(1+x^2)dt^2
\\ \label{backk2}
B_2 & = \frac{\varkappa \rho}{1+\varkappa^2(1+\rho^2)}d\psi_2 \wedge d\rho + \frac{\varkappa (1+\rho^2)^2 x }{1+\varkappa^2(1+\rho^2)^2(1+x^2)}d\psi_1 \wedge dx \ .
\end{align}

For the allowed ranges of $\rho$ and $x$ given in \eqref{ranges0} and \eqref{ranges12} the three metrics in \eqref{backk0}, \eqref{backk1} and \eqref{backk2} all have signature $(1,4)$. However, each describes a certain number of disconnected regions separated by curvature singularities. These regions can be classified by whether $t$ is a timelike or spacelike isometry and whether the NS-NS flux $H= dB$ is electric or magnetic:
\begin{itemize}
\item The metric \eqref{backk0} has two coordinate singularities, one at $\rho = \varkappa^{-1}$ and one at $\rho \to \infty$. Therefore, the space-time is split into two regions as shown in figure \ref{fig0}.
\item The metric \eqref{backk1} has two coordinate singularities, one at $x = \varkappa^{-1}(1+\rho^2)^{-1}$ and one at $\rho \to \infty$ or $x \to \infty$. Therefore, the space-time is again split into two regions as shown in figure \ref{fig1}.
\item The metric \eqref{backk2} has only one coordinate singularity at $\rho \to \infty$ or $x \to \infty$ and hence there is only a single region as shown in figure \ref{fig2}.
\end{itemize}
\begin{figure}[!p]
\begin{tikzpicture}[scale=1.75]
\fill[LightCyan2] (0,0) rectangle (3,1.5);
\fill[LavenderBlush2] (0,1.5) rectangle (3,3);
\fill[LavenderBlush3] (3,1.5) -- (3,3) -- (1.5,3);
\fill[LavenderBlush3] (0,3.1) -- (0,3.25) -- (3,3.25) -- (3,3.1);
\draw[draw=none,fill=LavenderBlush3] plot[domain=0.5:1] (3*\x,3/2*1/\x);
\draw[-,very thick] (0,0) -- (3,0);
\draw[-,very thick] (0,0) -- (0,3);
\draw[-,very thick,dotted,gray!75] (0,3) -- (0,3.1);
\draw[->,very thick] (0,3.1) -- (0,3.2);
\draw[-,very thick] (3,0) -- (3,3);
\draw[-,very thick,dotted,gray!75] (3,3) -- (3,3.1);
\draw[->,very thick] (3,3.1) -- (3,3.2);
\node[below] at (1.5,-0.05) {\small$x$};
\node[below] at (0,0) {\small$0$};
\node[below] at (3,0) {\small$1$};
\node[below left] at (0,3.2) {\small$\rho$};
\draw[gray!75,very thick,decoration={snake},decorate] (0,3.05)--(3,3.05);
\draw[-,dashed,very thick] (0,1.5) -- (3,1.5);
\draw[-,dashed,very thick] (0,3.25) -- (3,3.25);
\draw[-,very thick,gray!75] plot[domain=0.5:1] (3*\x,3/2*1/\x);
\node[left] at (0,1.5) {\small$\varkappa^{-1}$};
\node at (1,0.75) {\footnotesize$\mathbf{I}$};
\draw[very thick,black] (1,0.75) circle (0.2);
\node at (2,2.25) {\footnotesize$\mathbf{III}$};
\draw[very thick,black] (2,2.25) circle (0.2);
\node at (4,1.8) {\footnotesize$\mathbf{I}$};
\draw[very thick,black] (4,1.8) circle (0.2);
\node[right] at (4,1.8) {\ \ \, - timelike isometry and magnetic flux};
\node at (4,1.3) {\footnotesize$\mathbf{III}$};
\draw[very thick,black] (4,1.3) circle (0.2);
\node[right] at (4,1.3) {\ \ \, - spacelike isometry and electric flux};
\draw[very thick,fill=violet] (0,3.25) circle (0.05);
\draw[very thick,fill=violet!75] (9/2*1/2.75,2.75) circle (0.05);
\draw[very thick,fill=violet!50] (3,1.55) circle (0.05);
\draw[very thick,fill=violet!25] (1.5,1.55) circle (0.05);
\draw[very thick,fill=violet!0] (0,2.25) circle (0.05);
\draw[very thick,fill=cyan] (3,0.75) circle (0.05);
\draw[very thick,fill=cyan!75] (3,1.45) circle (0.05);
\draw[very thick,fill=cyan!50] (1.5,1.45) circle (0.05);
\draw[very thick,fill=cyan!25] (0,0.75) circle (0.05);
\draw[very thick,fill=cyan!0] (1.5,0) circle (0.05);
\end{tikzpicture}
\caption{Space-time structure for metric \eqref{backk0}.}\label{fig0}
\vspace{30pt}
\begin{tikzpicture}[scale=1.75]
\fill[Honeydew2] (0,0) rectangle (3,3);
\draw[draw=none,fill=LavenderBlush2] plot[domain=0:3] (\x,{1/(2/3)/sqrt(1+\x^2)}) -- (3,3) -- (0,3);
\draw[draw=none,fill=LavenderBlush1] plot[domain=0:3] (\x,{1/(2/3)/(1+\x^2)}) -- plot[domain=3:0] (\x,{1/(2/3)/sqrt(1+\x^2)});
\fill[LavenderBlush2] (0,3.1) -- (0,3.25) -- (3.25,3.25) -- (3.25,0) -- (3.1,0) -- (3.1,3.1);
\draw[-,very thick] (0,0) -- (3,0);
\draw[-,very thick,dotted,gray!75] (3,0) -- (3.1,0);
\draw[->,very thick] (3.1,0) -- (3.2,0);
\draw[-,very thick] (0,0) -- (0,3);
\draw[-,very thick,dotted,gray!75] (0,3) -- (0,3.1);
\draw[->,very thick] (0,3.1) -- (0,3.2);
\node[below left] at (3.2,0) {\small$\rho$};
\node[below left] at (0,3.2) {\small$x$};
\draw[gray!75,very thick,decoration={snake},decorate] (3.05,0)--(3.05,3.05);
\draw[gray!75,very thick,decoration={snake},decorate] (0,3.05)--(3.05,3.05);
\draw[-,dashed,very thick] (0,3.25) -- (3.25,3.25);
\draw[-,dashed,very thick] (3.25,0) -- (3.25,3.25);
\draw[-,dashed,very thick] plot[domain=0:3] (\x,{1/(2/3)/(1+\x^2)});
\draw[-,very thick,gray!75] plot[domain=0:3] (\x,{1/(2/3)/sqrt(1+\x^2)});
\node[left] at (0,1.5) {\small$\varkappa^{-1}$};
\node at (0.5,0.6) {\footnotesize$\mathbf{II}$};
\draw[very thick,black] (0.5,0.6) circle (0.2);
\node at (2.5,0.6) {\footnotesize$\mathbf{III}$};
\draw[very thick,black] (2.5,0.6) circle (0.2);
\node at (4,1.8) {\footnotesize$\mathbf{II}$};
\draw[very thick,black] (4,1.8) circle (0.2);
\node[right] at (4,1.8) {\ \ \, - timelike isometry and electric flux};
\node at (4,1.3) {\footnotesize$\mathbf{III}$};
\draw[very thick,black] (4,1.3) circle (0.2);
\node[right] at (4,1.3) {\ \ \, - spacelike isometry and electric flux};
\draw[very thick,fill=violet] (1.5,{1/(2/3)/sqrt(1+(3/2)^2)}) circle (0.05);
\draw[very thick,fill=violet!75] (3.25,0) circle (0.05);
\draw[very thick,fill=violet!50] (3.25,1.5) circle (0.05);
\draw[very thick,fill=violet!25] (1.5,3.25) circle (0.05);
\draw[very thick,fill=violet!0] (0,2.25) circle (0.05);
\end{tikzpicture}
\caption{Space-time structure for metric \eqref{backk1}.}\label{fig1}
\vspace{30pt}
\begin{tikzpicture}[scale=1.75]
\fill[LightCyan2] (0,0) rectangle (3,3);
\fill[LightCyan2] (0,3.1) -- (0,3.25) -- (3.25,3.25) -- (3.25,0) -- (3.1,0) -- (3.1,3.1);
\draw[-,very thick] (0,0) -- (3,0);
\draw[-,very thick,dotted,gray!75] (3,0) -- (3.1,0);
\draw[->,very thick] (3.1,0) -- (3.2,0);
\draw[-,very thick] (0,0) -- (0,3);
\draw[-,very thick,dotted,gray!75] (0,3) -- (0,3.1);
\draw[->,very thick] (0,3.1) -- (0,3.2);
\node[below left] at (3.2,0) {\small$\rho$};
\node[below left] at (0,3.2) {\small$x$};
\draw[gray!75,very thick,decoration={snake},decorate] (3.05,0)--(3.05,3.05);
\draw[gray!75,very thick,decoration={snake},decorate] (0,3.05)--(3.05,3.05);
\draw[-,dashed,very thick] (0,3.25) -- (3.25,3.25);
\draw[-,dashed,very thick] (3.25,0) -- (3.25,3.25);
\node at (1.5,1.5) {\footnotesize$\mathbf{I}$};
\draw[very thick,black] (1.5,1.5) circle (0.2);
\node at (4,1.55) {\footnotesize$\mathbf{I}$};
\draw[very thick,black] (4,1.55) circle (0.2);
\node[right] at (4,1.55) {\ \ \, - timelike isometry and magnetic flux};
\draw[very thick,fill=cyan] (1.5,0) circle (0.05);
\draw[very thick,fill=cyan!75] (3.25,1.5) circle (0.05);
\draw[very thick,fill=cyan!50] (1.5,3.25) circle (0.05);
\draw[very thick,fill=cyan!25] (0,1.5) circle (0.05);
\draw[very thick,fill=cyan!0] (0,0) circle (0.05);
\node[left] at (0,1.5) {\hphantom{\small$\varkappa^{-1}$}};
\end{tikzpicture}
\caption{Space-time structure for metric \eqref{backk2}.}\label{fig2}
\end{figure}

As the three backgrounds \eqref{backk0}, \eqref{backk1} and \eqref{backk2} are all meant to possess $q$-deformed $\SU(2,2)$ symmetry \cite{Delduc:2013fga} we may wonder if they are related. Indeed this is the case for $\AdS_3$, for which the isometry group is $\SO(2,2)$ and there are two inequivalent non-split $r$ matrices. It was shown in \cite{Hoare:2014oua} that the resulting two metrics are related by a $\varkappa$-dependent diffeomorphism. Motivated by this observation we look for diffeomorphisms between the three backgrounds \eqref{backk0}, \eqref{backk1} and \eqref{backk2}.\foot{\label{footac}While here we are interested in real diffeomorphisms, it is worth noting that the three backgrounds are formally related by the following $\varkappa$-independent analytic continuations
\begin{equation*}
t \to - \psi_1 \ , \quad \psi_1 \to -t \ , \quad \psi_2 \to \psi_2 \ , \quad \rho \to i\sqrt{1+\rho^2} \ , \quad x \to i x \ ,
\end{equation*}
which maps from \eqref{backk0} to \eqref{backk1} and
\begin{equation*}
t \to - \psi_2 \ , \quad \psi_1 \to \psi_1 \ , \quad \psi_2 \to - t \ , \quad \rho \to - i\sqrt{1+\rho^2} \ , \quad x \to \sqrt{1+x^2} \ ,
\end{equation*}
which maps from \eqref{backk0} to \eqref{backk2}.}
Indeed such transformations do exist. Firstly, using the coordinate transformation
\begin{equation}\begin{split}\label{coordtrans1}
\rho \to & \frac{\sqrt{-1+(1+\rho^2)(1+ x^2)}}{\sqrt{1+\varkappa ^2 (1+\rho^2) (1+x^2)}} \ , \qquad \quad \psi_1 \to -\psi_1 \ ,
\\
x \to & \frac{\rho \sqrt{1+\varkappa ^2 (1+\rho^2) (1+x^2)}}{\sqrt{1+\varkappa ^2 (1+\rho^2)}\sqrt{-1+(1+\rho^2)(1+x^2)}} \ ,
\end{split}\end{equation}
in \eqref{backk0} we find \eqref{backk2} up to a total derivative in the $B$-field. The coordinate ranges are mapped as follows
\begin{equation}\label{rangetrans1}
\rho \in [0,\varkappa^{-1}) \ , \quad x \in [0,1) \quad \to \quad \rho \in [0,\infty) \ , \quad x \in [0,\infty) \ ,
\end{equation}
and hence we find that region I in figures \ref{fig0} and \ref{fig2} are diffeomorphic. The coloured dots marking the boundaries of these regions indicate how they are mapped to each other. Secondly, using the coordinate transformation
\begin{equation}\begin{split}\label{coordtrans2}
\rho \to & \frac{\sqrt{1 + (1 + \rho^2)x^2 }}{\sqrt{-1 + \varkappa ^2 (1+ \rho^2)x^2 }} \ , \qquad \quad t \to \psi_2 \ , \quad \psi_1 \to t \ , \quad \psi_2 \to \psi_1 \ ,
\\
x \to & \frac{\rho \sqrt{-1 + \varkappa ^2 (1+ \rho^2)x^2 }}{\sqrt{1 + \varkappa ^2 (1+\rho^2)}\sqrt{1+(1+ \rho^2)x^2 }} \ ,
\end{split}\end{equation}
in \eqref{backk0} we find \eqref{backk1} again up to a total derivative in the $B$-field. This coordinate redefinition maps the coordinate ranges as follows
\begin{equation}\label{rangetrans2}
0 \leq x \leq \varkappa^{-1}\rho^{-1} \quad \text{for} \quad \rho \in (\varkappa^{-1},\infty) \quad \to \quad
 x \geq \varkappa^{-1}(1+\rho^2)^{-\frac{1}{2}} \quad \text{for} \quad \rho \in [0,\infty) \ ,
\end{equation}
Therefore we find that parts of region III in figures \ref{fig0} and \ref{fig1} are diffeomorphic. Again the coloured dots mark the boundaries of the regions that are related and indicate how they are mapped to each other. Note that in region III the three isometric directions $t$, $\psi_1$ and $\psi_2$ are all spacelike and hence we are free to interchange them as in the coordinate transformation \eqref{coordtrans2}.

Considering the deformed backgrounds as split up into regions separated by singularities, with the coordinates ranges inherited from the undeformed $\AdS_5$ geometry, we have seen that the three backgrounds \eqref{backk0}, \eqref{backk1} and \eqref{backk2} are all related by diffeomorphisms. However, there are still three distinct regions separated by the singularities. In all these regions the manifold has $(1,4)$ signature, but only for two of these is one of the $\U(1)$ isometries timelike. These two regions are further distinguished by the nature of the NS-NS flux -- in one it is magnetic (region I), while in the other it is electric (region II). The first of these regions has been explored in much detail and as such it is interesting to explore the differences between the two. In section \ref{sec:smat} we take a first step in this direction, computing the corresponding bosonic tree-level light-cone gauge $S$-matrices.

\subsection{Split deformations}\label{ssec:split}

\subsubsection*{r matrices}

Split $r$ matrices for $\mathfrak{su}(2,2)$ \cite{CahenGuttRawnsley} can be constructed from the canonical one \eqref{canonical} in a similar manner to inequivalent non-split $r$ matrices
\begin{equation}\label{rxrx}
R \to i R_X = i \Ad_X^{-1}  R \Adi_X \ , \qquad r \to i r_x = i (\Ad_X^{-1} \otimes \Ad_X^{-1})\, r \ .
\end{equation}
Here the factor of $i$ in \eqref{rxrx} means the resulting $r$ matrix satisfies the split mcYBe, and $X \in \SL(4;\mathbb{C})$ such that $iR_X$ preserves the defining relations \eqref{su22rep}. The existence of such elements is related to the non-compact nature of the algebra.

A practical way to obtain split solutions of the mcYBe is to analytically continue the non-split $r$ matrices \eqref{eq:non-splitrmatrices}, in the spirit of the discussion around \eqref{eq:su4rmat}. The procedure we follow is to exchange a pair of timelike and spacelike indices\foot{To recall, the timelike indices are $0$ and $5$, while $1$, $2$, $3$ and $4$ are spacelike. We could in fact exchange an odd number of pairs of timelike and spacelike indices, however as there are only two timelike indices any such permutation can be rewritten as first a permutation of spacelike indices and then an exchange of one pair of timelike and spacelike indices. Therefore, since permuting the spacelike indices of the non-split $r$ matrices \eqref{eq:non-splitrmatrices} maps them amongst themselves, here without loss of generality we can just consider the final exchange.} such that on each term an odd number of indices are flipped. As a result the $r$ matrices effectively pick up a factor of $i$ and hence become split $r$ matrices. Equivalently we can continue the non-split $\mathfrak{su}(4)$ $r$ matrix \eqref{eq:su4rmat} such that it picks up a factor of $i$. In the $\mathfrak{su}(4)$ picture there is manifestly only one way to do so up to inner automorphisms, resulting in
\begin{equation}\label{eq:splitrmatrices0}
r^{s} = m^{1}{}_{i}\wedge m^{i5} + m^{0}{}_{3}\wedge m^{32} + m^{0}{}_{4}\wedge m^{42} \ ,
\end{equation}
which is the result of exchanging indices $0$ and $1$ in $r_0$. As in the non-split case we have not proven that our procedure is exhaustive, though it seems unlikely that there are further split $r$ matrices. This $r$ matrix can be related to $R_0$ according to \eqref{rxrx}, with $R^s = i (R_0)_{P^s}$ and
\begin{equation}\label{p0s}
P^s = \exp\big[-\frac{i\pi}{2}m^{01}\big]\ ,
\end{equation}
where $R^s$ is the operator associated to $r^s$.

On the other hand, starting from the non-split $r$ matrices \eqref{eq:non-splitrmatrices} the method above gives a real result in two cases. The first is exchanging the indices $0$ and $1$ in $r_0$, and the second, exchanging the same pair of indices in $r_1$. For the former this gives \eqref{eq:splitrmatrices0}, while for the latter we find
\begin{align}
\label{eq:splitrmatrices1}
\tilde r^s & = m^{0}{}_{i} \wedge m^{i2} + m^{1}{}_{3} \wedge m^{35} + m^{1}{}_{4} \wedge m^{45} \ .
\end{align}
This $r$ matrix can be related to $R_0$ according to \eqref{rxrx}, with $\tilde R^s = i (R_0)_{\tilde P^s}$ and
\begin{equation}\label{p1s}
\tilde P^s = \exp\big[-\frac{i\pi}{2}(m^{02}-m^{15})\big]\exp\big[-\frac{i\pi}{2}m^{01}\big] = P_1 P^s \ ,
\end{equation}
where $\tilde R^s$ is the operator associated to $\tilde r^s$. Using the relation between $R_0$ and $R_1$ \eqref{p1p2}, this implies that $\tilde R^s = i (R_1)_{P^s}$. From \eqref{p0s} and \eqref{p1s} it then follows that the two split $r$ matrices \eqref{eq:splitrmatrices0} and \eqref{eq:splitrmatrices1} are related by an $\SU(2,2)$ transformation
\begin{equation}\label{q}
\tilde R^{s} = \Ad^{-1}_{Q} R^{s} \Adi_{Q} \ , \qquad Q = (P^s)^{-1} \tilde P^s = (P^s)^{-1} P_1 P^s = \exp[-\frac{\pi}{2}(m^{05}-m^{12})] \ ,
\end{equation}
and hence, in agreement with the analysis above, are not inequivalent.

Before we proceed to explore the effect of the deformation associated to the split $r$ matrix, let us note that the key difference between the non-split and split $r$ matrices is the real form of the Cartan subalgebra with which they commute. The split $r$ matrix commutes with the generators of $\mathbb{R}^2 \times \U(1) \subset \mathrm{SU}(2,2)$, in contrast to the non-split ones which commute with the generators of $\U(1)^3$.

\subsubsection*{Geometries}

For the backgrounds following from the split $r$ matrix \eqref{eq:splitrmatrices0} it is natural to choose coordinates that manifest the $\mathbb{R}^2 \times \U(1)$ symmetry preserved by the deformations. As a result, in the $\sdp \to 0$ limit, they will only cover a part of $\AdS_5$. Therefore, we may consider the effect of the deformation on various regions of the space. In general there are a number of choices. Here we present four backgrounds that are related to the non-split ones by analytic continuation. Note that in this section we will restrict our discussion of the metrics to the regions that recover $\AdS_5$ in the $\sdp \to 0$ limit, i.e. we will not consider the space-times beyond singularities.

The first coset parametrisation we consider is
\begin{equation}
g^s_{0,\lambda} = \lambda^s e^{\arcsin{x} \, m^{23}} e^{\arcsinh{\rho} \, m^{25}} \ , \qquad \rho \in (-\infty,\infty) \ , \quad x \in [0,1) \ ,
\end{equation}
with
\begin{equation}\label{lst}
\lambda^s = e^{\psi_1 m^{15} - t m^{02} - \psi_2 m^{34}} \ .
\end{equation}
The background for the model based on $(r^s,g^s_{0,\lambda})$ is
\begin{align}
\nonumber
ds^2 & = \frac1{1+\sdp^2\rho^2} \Big((1+\rho^2)d\psi_1^2 + \frac{d\rho^2}{1+\rho^2}\Big) + \frac1{1-\sdp^2\rho^4x^2}\Big(-\rho^2(1-x^2)dt^2 +\frac{\rho^2 dx^2}{1-x^2}\Big) + \rho^2 x^2 d\psi_2^2 \ ,
\\ \label{backsplittime0}
B & = \frac{\sdp\rho}{1+\sdp^2\rho^2} d\psi_1 \wedge d\rho + \frac{\sdp \rho^4 x}{1-\sdp^2 \rho^4 x^2} dt \wedge dx \ ,
\end{align}
where we restrict to the region $x < \sdp^{-1}\rho^{-2}$. As for the $r$ matrix, the deformed geometry can be understood as an analytic continuation of the non-split background \eqref{backk0}. Indeed, replacing $(t,\psi_1,\varkappa) \to (-i\psi_1,-it,i\sdp)$ in \eqref{backk0} reproduces \eqref{backsplittime0}.\label{pageac1}

The second coset parametrisation we choose is
\begin{equation}\label{finalcp}
g_{1,\lambda}^s = Q \lambda^s e^{\arcsinh{x} \, m^{35}}e^{\arcsinh{\rho} \, m^{25}} \ , \qquad \rho \in (-\infty,\infty) \ , \quad x \in [0,\infty) \ ,
\end{equation}
where $Q$ is given in \eqref{q}.
In this case the background for $(r^s,g^s_{1,\lambda})$ is
\begin{align} \nonumber
ds^2 & = \frac1{1-\sdp^2(1+\rho^2)}\Big(- \rho^2 dt^2 + \frac{d\rho^2}{1+\rho^2}\Big)
\\ \nonumber & \quad + \frac1{1+\sdp^2(1+\rho^2)^2x^2}\Big((1+\rho^2)(1+x^2)d\psi_1^2 + \frac{(1+\rho^2)dx^2}{1+x^2} \Big) + (1+\rho^2)x^2d\psi_2^2
\\\label{backsplittime1}
B & = \frac{\sdp\rho}{1-\sdp^2(1+\rho^2)} dt \wedge d\rho + \frac{\sdp(1+\rho^2)^2x}{1+\sdp^2(1+\rho^2)^2x^2}d\psi_1\wedge dx \ ,
\end{align}
where we restrict to the region $\rho < \sdp^{-1}(1-\sdp^2)^{\frac12}$ and for $t$ to remain a timelike isometry we require that $\sdp < 1$. Due to the constant factor of $Q$ in the coset parametrisation \eqref{finalcp}, which is the same group element that related $R^s$ and $\tilde R^s$ \eqref{q}, we expect that this deformed geometry is related to the non-split background \eqref{backk1} by analytic continuation. Indeed, setting $(t,\psi_1,\varkappa) \to (-i\psi_1,-it,i\sdp)$ in \eqref{backk1} reproduces \eqref{backsplittime1}.

The third coset parametrisation we take is
\begin{equation}
g^s_{0,\tl} = \tl^s e^{\arcsinh{x} \, m^{03}} e^{\arcsin{t} \, m^{05}} \ , \qquad t \in (-1,1) \ , \quad x \in [0,\infty) \ ,
\end{equation}
with
\begin{equation}\label{lss}
\tl^s = e^{\psi_3 m^{15} - \psi_1 m^{02} - \psi_2 m^{34}} \ ,
\end{equation}
The background for the model based on $(r^s,g^s_{0,\tl})$ is
\begin{align}
\nonumber
ds^2 & = \frac1{1-\sdp^2t^2}\Big((1-t^2)d\psi_3^2 - \frac{dt^2}{1-t^2}\Big)
         + \frac1{1+\sdp^2 t^4 x^2} \Big( t^2(1+x^2) d\psi_1^2 + \frac{t^2 dx^2}{1+x^2}\Big) + t^2 x^2 d\psi_2^2 \\
B & = - \frac{\sdp t}{1-\sdp^2 t^2} d\psi_3 \wedge dt - \frac{\sdp t^4x}{1+\sdp^2 t^4x^2} d\psi_1 \wedge dx \ ,
\label{backsplitspace0}
\end{align}
where we consider the region $-\sdp^{-1} < t < \sdp^{-1}$. Note that initially we had $t \in (-1,1)$ and hence if $\sdp < 1$ there is no additional restriction on the range of $t$. This background is reproduced from \eqref{backk0} by the analytic continuation $(t,\psi_1,\rho,x,\varkappa) \to (-i\psi_3,-i\psi_1,-it,-ix,i\sdp)$.

The final coset parametrisation that we present here is
\begin{equation}
g^s_{1,\tl} = \tl^s e^{\arcsinh{t} \, m^{03}} e^{\arcsinh{\rho} \, m^{35}} \ , \qquad \rho \in (-\infty,\infty) \ , \quad t \in [0,\infty) \ ,
\end{equation}
In this case the background for $(r^s,g^s_{1,\tl})$ is
\begin{align}
\nonumber
ds^2 & = \frac1{1+\sdp^2\rho^2}\Big((1+\rho^2)d\psi_3^2 + \frac{d\rho^2}{1+\rho^2}\Big)
         + \frac1{1-\sdp^2 \rho^4 (1+t^2)} \Big( \rho^2t^2 d\psi_1^2 - \frac{\rho^2 dt^2}{1+t^2}\Big) + \rho^2(1+t^2) d\psi_2^2 \\
B & = \frac{\sdp \rho}{1+\sdp^2 \rho^2} d\psi_3 \wedge d\rho + \frac{\sdp \rho^4 t}{1-\sdp^2 \rho^4 (1+t^2)} d\psi_1 \wedge dt \ ,
\label{backsplitspace1}
\end{align}
where we restrict to the region $t < \sdp^{-1}\rho^{-2}(1-\sdp^2\rho^4)^{\frac12}$. This background is reproduced from \eqref{backk0} by the analytic continuation $(t,\psi_1,\rho,x,\varkappa) \to (-i\psi_3,-i\psi_1,-\rho,\sqrt{1+t^2},i\sdp)$.\label{pageac2}

A few comments are in order. The backgrounds \eqref{backsplittime0} and \eqref{backsplittime1} both have one timelike and two spacelike isometries. On the other hand backgrounds \eqref{backsplitspace0} and \eqref{backsplitspace1} have three spacelike isometries. However, these pairs cannot be diffeomorphic as in each case the former has electric NS-NS flux, $H=dB$, while the latter has magnetic. Finally let us note that, as for the non-split deformations, these backgrounds all have curvature singularities for certain values of the non-isometric coordinates.

In the $\sdp \to 0$ limit these metrics cover different regions of $\AdS_5$. To analyse this it is useful to introduce embedding coordinates
\begin{equation}\label{embbbcon}
-Z_0^2 - Z_5^2 + \sum_{i=1}^4 Z_i^2 = - 1  \ .
\end{equation}
For the background \eqref{backsplittime0} the coordinate patch of $\mathrm{AdS}_5$ can be obtained from the parametrisation
\begin{align}
\nonumber
& Z_0 \pm Z_1 = \sqrt{1+\rho^2} \, e^{\pm \psi_1}  \ , \qquad Z_2 \pm Z_5 = \rho\sqrt{1-x^2} \, e^{\pm t} \ ,  \qquad Z_3 \pm i Z_4 = \rho x \, e^{\pm i \psi_2} \ ,\\
& Z_0 > 0 \ , \qquad Z_0^2 > Z_1^2 \ , \qquad Z_5^2 < Z_2^2 \ ,
\end{align}
while for \eqref{backsplittime1} we can take
\begin{align}
\nonumber
& Z_1 \pm Z_0 = \rho \, e^{\pm t} \ , \qquad Z_5 \pm Z_2 = \sqrt{1+\rho^2}\sqrt{1+x^2} \, e^{\pm\psi_1} \ , \qquad  Z_3 \pm i Z_4  = \sqrt{1+\rho^2} x \, e^{\pm i \psi_2} \ ,\\
& Z_5 > 0 \ , \qquad Z_0^2 < Z_1^2 \ , \qquad Z_5^2 > Z_2^2 \ .
\end{align}
For \eqref{backsplitspace0} we can use
\begin{align}
\nonumber
& Z_0 \pm Z_1 = \sqrt{1-t^2} \, e^{\pm \psi_3} \ , \qquad  Z_5 \pm Z_2  = t\sqrt{1+x^2} \, e^{\pm \psi_1} \ , \qquad  Z_3 \pm i Z_4  = t x \, e^{\pm i \psi_2} \ ,\\
& Z_0 > 0 \ , \qquad Z_0^2 > Z_1^2 \ , \qquad Z_5^2 > Z_2^2 \ , \qquad Z_5^2 -Z_2^2 > Z_3^2+Z_4^2 \ ,
\end{align}
and finally for \eqref{backsplitspace1} we can set
\begin{align}
\nonumber
& Z_0 \pm Z_1 = \sqrt{1+\rho^2} \, e^{\pm \psi_3} \ , \qquad  Z_5 \pm Z_2  = \rho t \, e^{\pm \psi_1} \ , \qquad  Z_3 \pm i Z_4  = \rho\sqrt{1+t^2} \, e^{\pm i \psi_2} \ ,\\ &
Z_0 > 0 \ , \qquad Z_0^2 > Z_1^2 \ , \qquad Z_5^2 > Z_2^2 \ , \qquad Z_5^2 -Z_2^2 < Z_3^2+Z_4^2 \ .
\end{align}
In each of these expressions the inequalities govern the part of $\AdS_5$ that is covered by the parametrisation. Note that $\AdS_5$ does not have a region with $Z_0^2 < Z_1^2$ and $Z_5^2 < Z_2^2$ as these inequalities are incompatible with the embedding constraint \eqref{embbbcon}.

\begin{figure}[t]
\begin{center}
\includegraphics{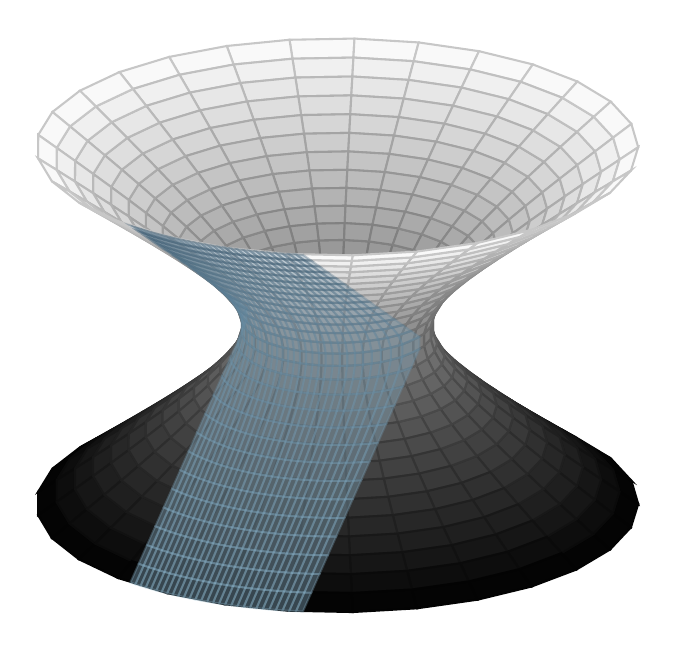}
\caption{$\AdS_2$ represented as a hyperboloid, illustrating the region covered by the parametrization \eqref{ads2ss}.}\label{fig:splitcoordcover}
\end{center}
\end{figure}

Truncating the third case to an $\AdS_2$ subspace
\begin{equation}\label{ads2ss}
Z_0 \pm Z_1 = \sqrt{1-t^2} \, e^{\pm \psi_3} \ , \qquad Z_5 = t \ ,
\end{equation}
figure \ref{fig:splitcoordcover} illustrates the region covered by these coordinates, namely a subregion of the Poincar\'e patch.
While we will not give the explicit metrics and $B$-fields as they are not particularly illuminating, it is interesting to note that when considering the deformation of the Poincar\'e patch of $\AdS_5$ in some sense it is natural to consider the split $r$ matrix as opposed to the non-split one. The reason for this is that the manifest symmetry of the Poincar\'e patch is $\mathbb{R} \ltimes \mathrm{ISO}(1,3)$, where the first factor is the scaling symmetry and the second factor is the Poincar\'e group. This has an $\mathbb{R}^2 \times \U(1)$ subgroup consisting of the scaling symmetry and a commuting boost and rotation from the Lorentz subgroup $\SO(1,3)$. This is precisely the isometry of $\AdS_5$ preserved by the split $r$ matrix. On the other hand the isometry preserved by the non-split $r$ matrix is $\U(1)^3$, which is not a subgroup of the symmetries of the Poincar\'e patch.

\subsection{Contraction limits}\label{ssec:contractions}

The non-split and split deformations of $\AdS_5$ discussed above admit certain contraction limits, named for their relation to contractions of the symmetry algebra. They are generalisations of the flat-space limit of the undeformed $\AdS_5$ model, in which case the isometry algebra of $\AdS_5$ is contracted to the 5-d Poincar\'e algebra
\begin{equation}
\mathfrak{su}(2,2) \simeq \mathfrak{so}(2,4) \to \mathfrak{iso}(1,4) \ .
\end{equation}

The contractions we consider here are of the type discussed in \cite{Pachol:2015mfa}. Considering the generators $m^{ij}$ we select one index $\hat \imath$ and scale those generators containing that index to infinity
\begin{equation}\label{mscale}
m^{i\hat \imath} \to \scale m^{i\hat \imath} \ , \qquad  \scale \to \infty \ .
\end{equation}
For the undeformed algebra $\mathfrak{su}(2,2) \simeq \mathfrak{so}(2,4)$ this gives $\mathfrak{iso}(1,4)$ if $\hat \imath$ is a timelike index ($0$ or $5$), or alternatively $\mathfrak{iso}(2,3)$ if it is spacelike ($1,\ldots,4$). The contractions continue to be admissible in the $q$-deformed algebra so long as we additionally scale $q$ as $\log q \to - \scale^{-1} \log q$.\foot{For a well-defined contraction limit we require that $\log q \to - \scale^{-\alpha} \log q$ with $\alpha \geq 1$. However it is only for $\alpha = 1$ that the algebra remains non-trivially deformed.} In the first case we find $\mathcal{U}_{\kappa}(\mathfrak{iso}(1,4))$, a deformation of the 5-d Poincar\'e algebra, and in the second case $\mathcal{U}_{\kappa}(\mathfrak{iso}(2,3))$. Such deformations have been extensively studied in the literature and are referred to as $\kappa$-Poincar\'e algebras \cite{Lukierski:1991pn}. Depending on whether we contract an algebra with $q$ real or $q$ a phase factor, they are of so-called timelike or spacelike type, see, for example, \cite{Borowiec:2013lca} for a unified discussion of $\kappa$-Poincar\'e algebras for any dimension and signature.

In the undeformed model the contraction is manifestly compatible with the action \eqref{def:non-split} if the selected index is $5$. This is because the generators $m^{i5}$ form a basis of the space onto which the projector $P$ projects. To preserve the finiteness of the bilinear form on this space we also also need to scale $\Tr \to \scale^{-2} \Tr$ and correspondingly scale the effective string tension $T \to \scale^2 T$. The physical interpretation of this is the restoration of the $\AdS_5$ radius and that the flat-space limit amounts to taking this radius to infinity.

These observations continue to be true in the deformed model, however we additionally require that the $r$ matrix is compatible with the contraction limit. To understand the implications of this we start by recalling that in the non-split case the $q$-deformed symmetry of the model has \cite{Delduc:2013fga}
\begin{equation}
q = \exp\big[-\frac{\varkappa}{T}\big]\ ,
\end{equation}
that is $q$ is real. While in the split case we have that $q$ is a phase factor
\begin{equation}
q = \exp\big[-\frac{i\sdp}{T}\big] \ ,
\end{equation}
by analytic continuation.
As $\log q \sim \scale^{-1}$ to retain a finite deformation and $T \sim \scale^2$, we must have that $\varkappa,\sdp \sim \scale$. Therefore, for the combination $\varkappa R$ in the action \eqref{def:non-split} to also remain finite the index $5$ should occur a maximum of one time in each term of the corresponding $r$ matrix \cite{Pachol:2015mfa}.\foot{Recall that in the definition of the $R$ operator in terms of the $r$ matrix \eqref{Rrmap} there is a single instance of the bilinear form, which scales like $\scale^{-2}$.} This is the case for the $r$ matrices $r_0$ and $r^s$ in \eqref{eq:non-splitrmatrices} and \eqref{eq:splitrmatrices0} respectively. After taking the limit these $r$ matrices can be be understood as solutions of the mcYBe \eqref{rrmcybe} for the Poincar\'e algebra $\mathfrak{iso}(1,4)$ -- the so-called timelike $\kappa$-Poincar\'e $r$ matrix in the non-split case, and the spacelike one in the split case. In this way they can be used to directly construct the deformations of flat space \cite{Borowiec:2015wua} that follow from contraction limits \cite{Pachol:2015mfa}.

\

To translate these contractions into limits of the backgrounds we look at the coset parametrisations and read off which fields should be rescaled by $\scale^{-1}$ to maintain finiteness of the group element. We additionally rescale the metric and $B$-field by $\scale^2$ (from the rescaling of the effective string tension $T$), set $\varkappa = \scale \kappa^{-1}$ in the non-split and $\sdp = \scale \sdpc^{-1}$ in the split backgrounds and take $\scale \to \infty$. Following this procedure we find the following contracted backgrounds.\foot{Note that even though the background \eqref{backsplittime1} is based on $r^s$ it does not have a good contraction limit due to the constant factor of $Q$ (which depends on $m^{05}$) in the coset parametrisation \eqref{finalcp}. Furthermore, the limit is not compatible with the restriction $\sdp < 1$ required for $t$ to remain a timelike isometry of the background.}
\begin{itemize}
\item For the background \eqref{backk0} based on $(r_0,g_0)$ we rescale $t \to \scale^{-1} t$ and $\rho \to \scale^{-1}\rho$. Taking $\scale \to \infty$ we find
\begin{equation}\begin{split}\label{conbackns0}
ds^2 & = \frac1{1-\kappa^{-2}\rho^2}\big( - dt^2 + d\rho^2 \big)
          + \rho^2\big( (1-x^2) d\psi_1^2 + \frac{dx^2}{1-x^2} +  x^2 d\psi_2^2\big) \ ,
\\ B & = \frac{\kappa^{-1} \rho}{1-\kappa^{-2}\rho^2}dt \wedge d\rho \ .
\end{split}\end{equation}
\item For the background \eqref{backsplittime0} based on $(r^s,g_{0,\lambda}^s)$ we rescale $\psi_1 \to \scale^{-1} \psi_1$ and $\rho \to \scale^{-1} \rho$ to give
\begin{equation}\begin{split}\label{con1}
ds^2 & = \frac1{1+\sdpc^{-2}\rho^2} \big(d\psi_1^2 + d\rho^2\big) + \rho^2 \big(-(1-x^2)dt^2 +\frac{dx^2}{1-x^2} + x^2 d\psi_2^2\big) \ ,
\quad
\\ B & = \frac{\sdpc^{-1}\rho}{1+\sdpc^{-2}\rho^2}d\psi_1\wedge d\rho \ .
\end{split}\end{equation}
\item For the background \eqref{backsplitspace0} based on $(r^s,g_{0,\tilde\lambda}^s)$ we rescale $\psi_3 \to \scale^{-1} \psi_3$ and $t\to \scale^{-1} t$ to give
\begin{equation}\begin{split}\label{con2}
ds^2 & = \frac1{1-\sdpc^{-2}t^2}\big(d\psi_3^2 - dt^2\big) + t^2 \big( (1+x^2) d\psi_1^2 + \frac{dx^2}{1+x^2} + x^2 d\psi_2^2\big) \ ,
\quad
\\ B & = -\frac{\sdpc^{-1} t}{1-\sdpc^{-2} t^2} d\psi_3\wedge dt \ .
\end{split}\end{equation}
\item For the background \eqref{backsplitspace1} based on $(r^s,g_{1,\tilde \lambda}^s)$ we rescale $\psi_3 \to \scale^{-1} \psi_3$ and $\rho \to \scale^{-1} \rho$ to give
\begin{equation}\begin{split}\label{con3}
ds^2 & = \frac1{1+\sdpc^{-2}\rho^2}\big(d\psi_3^2 + d\rho^2\big) + \rho^2\big( t^2 d\psi_1^2 - \frac{ dt^2}{1+t^2} + (1+t^2) d\psi_2^2 \big) \ ,
\quad
\\ B & = \frac{\sdpc^{-1}\rho}{1+\sdpc^{-2}\rho^2}d\psi_3\wedge d\rho \ .
\end{split}\end{equation}
\end{itemize}
A few comments are in order. The first of these four cases follows from a non-split $r$ matrix and as such has timelike $\mathcal{U}_\kappa(\mathfrak{iso}(1,4))$ symmetry. This limit was first studied in \cite{Arutynov:2014ota} and is such that one stays in region I in figure \ref{fig0}. The resulting background is known as the mirror background and is related to the mirror $\AdS_5 \times S^5$ superstring \cite{Arutynov:2014ota,Arutyunov:2014cra,Arutyunov:2014jfa}. In the remaining three cases the starting point is the split $r$ matrix. These cases have spacelike $\mathcal{U}_{\sdpc} (\mathfrak{iso}(1,4))$ symmetry.

In all cases the $B$-field becomes a total derivative in the limit. Dropping the $B$-field in \eqref{conbackns0} and T-dualising in $t$ we find the metric for 5-d de-Sitter space, $dS_5$, in static coordinates \cite{Arutynov:2014ota}. On the other hand, for the split cases, dropping the $B$-field and T-dualising in $\psi_1$ in \eqref{con1} and $\psi_3$ in \eqref{con2} and \eqref{con3}, we find metrics for $\AdS_5$. After dualisation the parameter $\kappa$ corresponds to the radius of the symmetric space. Indeed in all four metrics, by rescaling the two distinguished coordinates, one can arrange for $\kappa$ to only appear in an overall factor of $\kappa^2$. Finally, let us note that these contraction limits all commute with the limit in which the deformation parameter ($\varkappa$ or $\sdp$) goes to zero -- taking $\kappa \to \infty$ in the above metrics gives various forms of the flat space metric.

\

We conclude this section with a curious observation. There do exist contraction limits for which the selected index in \eqref{mscale} is different from $5$. After dropping the total derivative $B$-field, these contractions correspond to finite and non-degenerate limits of the background. While they are sensible at the level of the algebra, the limits appear to be ill-defined in the action \eqref{def:non-split} and $r$ matrices.

The cases for which this works all have the same structure. In the coset elements there is, by construction, one isometry whose corresponding generator has an index not shared with either of the generators associated to the non-isometric coordinates. This isometry does not appear in the $B$-field and its metric component is independent of the deformation parameter. For the backgrounds computed in sections \ref{ssec:non-split} and \ref{ssec:split} the associated contraction always gives a finite and non-degenerate metric and an infinite total derivative $B$-field. Unlike the well-defined contraction limits discussed above, these limits do not commute with taking the deformation parameter ($\varkappa$ or $\sdp$) to zero. Indeed, without additionally rescaling certain coordinates by $\kappa$ or $\kappa^{-1}$, taking $\kappa \to \infty$ does not give a finite result.

While it is not clear why this construction should work, let us briefly discuss these limits.
\begin{itemize}
\item Contracting the background \eqref{backk2} based on $(r_2,g_2)$ by selecting index $0$, we rescale $t \to \scale^{-1} t$. Taking $\scale \to \infty$, redefining $\rho \to \frac{\rho x}{\sqrt{1-\rho^2 x^2}}$, $x \to \frac{\rho \sqrt{1-x^2}}{\sqrt{1-\rho^2}}$ and then rescaling $\rho \to \kappa^{-1}\rho$, we find \eqref{conbackns0} up to an infinite total derivative $B$-field. As discussed above, T-dualising in $t$ we find $dS_5$ in static coordinates with $\kappa$ corresponding to the radius.
\item Contracting the background \eqref{backk0} based on $(r_0,g_0)$ by selecting index $4$, we rescale $\psi_2 \to \scale^{-1} \psi_2$. Additionally redefining $\rho \to \rho^{-1}$ and $x \to x^{-1}$ and dropping the infinite total derivative $B$-field we find
\begin{align}\label{thisone}
ds^2 & = \kappa^{2}\big((1+\rho^2) dt^2 - \frac{d\rho^2}{1+\rho^2} \big) + \kappa^2\rho^2\big( (x^2-1) d\psi_1^2 + \frac{dx^2}{x^2-1}\big) + \frac{d\psi_2^2}{\rho^2x^2} \ .
\end{align}
This is a limit in which one stays in region III in figure \ref{fig0} and was first considered in \cite{Hoare:2014pna}, in which it was noted that T-dualising in $\psi_2$ also gives a metric for $dS_5$ with radius $\kappa$.\foot{Rescaling $\psi_2 \to \kappa \psi_2$, $\kappa$ only appears in an overall factor of $\kappa^2$.} This metric covers the complementary patch to that covered by static coordinates. The metric \eqref{thisone} can also be found as a contraction of \eqref{backk1} again selecting index $4$. This corresponds to rescaling $\psi_2 \to \scale^{-1} \psi_2$, and, after taking the limit, redefining $\rho \to \frac{\rho\sqrt{x^2-1}}{\sqrt{1-\rho^2(x^2-1)}}$, $x\to \frac{\sqrt{1-\rho^2(x^2-1)}}{\rho x}$.
\item Finally, selecting index $4$ the corresponding contraction limit ($\psi_2 \to \scale^{-1} \psi_2$) can be considered in all four backgrounds \eqref{backsplittime0}, \eqref{backsplittime1},\foot{Note that here the index $4$ does not appear in the constant factor $Q$ in the coset parametrisation \eqref{finalcp}.} \eqref{backsplitspace0}, \eqref{backsplitspace1} based on the split $r$ matrix $r^s$. Taking these limits requires formally extending outside the allowed ranges of the coordinates or the parameter $\sdp$. While we do not give the backgrounds explicitly they again all have an infinite total derivative $B$-field. As for the split cases above they are also all T-dual to $\AdS_5$ with radius $\sdpc$.
\end{itemize}

\subsection{T-dual supergravity backgrounds by analytic continuation}\label{ssec:sugra}
The deformations of $\AdS_5$ discussed in sections \ref{ssec:non-split} and \ref{ssec:split} are based on solutions of the mcYBe over $\mathfrak{su}(2,2)$. Whether these deformations can be extended to the full $\AdS_5 \times S^5$ superstring depends on if the corresponding solution of the mcYBe can be extended to the superalgebra $\mathfrak{psu}(2,2|4)$ \cite{Delduc:2013qra,Delduc:2014kha}. This is possible for the three non-split cases, but not for the split case. For the split $r$ matrix there will be an obstruction due to the compact $\mathfrak{su}(4)$ subalgebra, for which it is known there are no split solutions. Indeed, starting from the standard non-split $r$ matrix for $\mathfrak{su}(4)$, there is no $X \in \SL(4;\mathbb{C})$ such that $iR_X$ in \eqref{rxrx} preserves the compact real form. It is also worth noting that each of the non-split $r$ matrices potentially has a number of different inequivalent extensions to the superalgebra, for example corresponding to the various Dynkin diagrams of $\mathfrak{psu}(2,2|4)$.

The $r$ matrices we have studied above are all related to \eqref{canonical} by the action of a constant element of $\SL(4;\mathbb{C})$ \eqref{rxrxns} (and multiplication by $i$ in the split case \eqref{rxrx}). One way to extend the relevant matrices, given in \eqref{p1p2} and \eqref{p0s}, to $\PSL(4|4;\mathbb{C})$ is
\begin{equation}\label{pidrot}
\begin{pmatrix} P & 0 \\ 0 & \mathbf{1}_{4\times 4} \end{pmatrix} \ .
\end{equation}
We can then act with these on the standard non-split $\mathfrak{psu}(2,2|4)$ $r$ matrix to give new $\mathfrak{psl}(4|4;\mathbb{C})$ $r$ matrices.

In the non-split cases the new $r$ matrices preserve the real form of the superalgebra.\foot{We define the real form $\mathfrak{su}(2,2|4)$ following the conventions of \cite{Arutyunov:2009ga}. It is given by the set of elements in the $8 \times 8$ matrix representation of $\mathfrak{sl}(4|4;\mathbb{C})$ satisfying $M^\dagger \Sigma + \Sigma M = 0$, where $\Sigma = \diag(1,1,-1,-1,1,1,1,1)$. The superalgebra $\mathfrak{psu}(2,2|4)$ is then given by modding out by the central element.} As the deformed model \eqref{def:non-split} based on the pair $(R_X,g)$ is equivalent to that based on $(R,Xg)$, we can conjecture that the backgrounds following from the new $r$ matrices are $\varkappa$-independent analytic continuations (such as those given in footnote \ref{footac}) of the background following from the original $r$ matrix. As the new $r$ matrices preserve the real form of the full superalgebra $\mathfrak{psu}(2,2|4)$ the resulting backgrounds should also have the correct reality properties.

In the split case the new $r$ matrix preserves the real form of the $\mathfrak{su}(2,2)$ subalgebra, but not of the $\mathfrak{su}(4)$ subalgebra due to the factor of $i$ in \eqref{rxrx}. However, we can still formally use that the deformed model \eqref{def:non-split} based on $(iR_X,g)$ is equivalent to that based on $(R,Xg)$, with $\varkappa = i\sdp$. This allows us to again conjecture that the background following from the new $r$ matrix is an analytic continuation of the original background, along with the replacement $\varkappa \to i\sdp$ (for example, those transformations given on pages \pageref{pageac1} and \pageref{pageac2}). As the new $r$ matrix does not preserve the real form of the bosonic subalgebra, let alone the full superalgebra, we do not expect the resulting background to have the correct reality properties.

In table \ref{tab:ac} we have listed the candidate analytic continuations (i.e. those that give the various deformations of $\AdS_5$ and are of the required form) that we have already encountered.
\begin{table}[t]
\begingroup
\renewcommand{\arraystretch}{1.6}
\begin{tabular}{l|l|l|l}
& Background map & $r$ matrix map & \parbox[top][1.5cm][c]{4cm}{Analytic continuation \\ $(t,\psi_1,\psi_2,\rho,x) \to$} \\
\hline \hline
{\em Non-split} & \eqref{backk0} $\to$ \eqref{backk1} & $R_0 \to R_1 = (R_0)_{P_1} $ & $(-\psi_1,-t,\psi_2,i\sqrt{1+\rho^2},i x)$
\\
& \eqref{backk0} $\to$ \eqref{backk2} & $R_0 \to R_2 = (R_0)_{P_2} $ & $(-\psi_2,\psi_1,-t,-i\sqrt{1+\rho^2},\sqrt{1+x^2})$
\\ \hdashline
{\em Split}, $\varkappa \to i\sdp$ & \eqref{backk0} $\to$ \eqref{backsplittime0} & $R_0 \to R^s_0 = (R_0)_{P^s}$ & $(-i\psi_1,-it,\psi_2,\rho,x)$
\\
& \eqref{backk0} $\to$ \eqref{backsplittime1} & $R_0 \to R^s_0 = (R_0)_{\hat P^s}$ & $(it,i\psi_1,\psi_2,i\sqrt{1+\rho^2},- i x)$
\\
& \eqref{backk0} $\to$ \eqref{backsplitspace0} & $R_0 \to R^s_0 = (R_0)_{P^s}$ & $(-i\psi_3,-i\psi_1,\psi_2,-it,-ix)$
\\
& \eqref{backk0} $\to$ \eqref{backsplitspace1} & $R_0 \to R^s_0 = (R_0)_{P^s}$ & $(-i\psi_3,-i\psi_1,\psi_2,-\rho,\sqrt{1+t^2})$
\end{tabular}
\endgroup
\caption{Analytic continuations between the non-split and split deformations of $\AdS_5$.}\label{tab:ac}
\end{table}
For now just considering the deformed $\AdS_5$ theories, the analytic continuations are equivalent to using the new $r$ matrices if
\begin{equation}\label{chain}
(R_0,g_0^{a.c}) \sim (R_0, P_1 g_1) \sim ((R_0)_{P_1} , g_1) \sim (R_1,g_1) \ ,
\end{equation}
where by $\sim$ we mean the two corresponding backgrounds are equal. Similar chains hold for the remaining cases. The latter two steps are identities, and therefore it remains to check the first step. The condition we require is
\begin{equation}\label{accond}
g_0^{a.c.} = P_1 g_1 h \ , \qquad h \in \SO(1,4;\mathbb{C}) \ ,
\end{equation}
that is $g_0^{a.c.}$ and $P_1 g_1$ are related by a complexified gauge transformation. If this is the case then by gauge invariance $h$ will drop out of the action and we will have the desired equivalence. For the first, second, fifth and sixth rows in table \ref{tab:ac}, \eqref{accond} indeed holds for the group elements used in sections \ref{ssec:non-split} and \ref{ssec:split}.

For the third and fourth rows it is useful to introduce the alternative coset parametrisations
\begin{equation}\begin{split}
\hat g_0 & = \lambda e^{ \arcsin{x} \, m^{23}} e^{ \arcsinh{\rho}\, m^{25}} \ ,
\\
\hat g_1 & = \lambda e^{\arcsinh x \, m^{35}} e^{\arcsinh\rho \, m^{25}} \ ,
\end{split}\end{equation}
which are related to \eqref{eq:gns0} by interchanging indices $1$ and $2$. The backgrounds associated to $(R_0,\hat g_0)$ and $(R_1,\hat g_1)$ are given by \eqref{backk0} and \eqref{backk1} respectively as expected.
Furthermore, $\hat g_0$ and $\hat g_1$ are also related by analytic continuation,
\begin{equation}\label{g0acg1}
\hat g_0^{a.c.} = \hat P_1 \hat g_1 h \ , \qquad h \in \SO(1,4;\mathbb{C}) \ ,
\end{equation}
where $\hat P_1 = i \exp[-\frac{i\pi}{2}(m^{01} + m^{25})]$ is an alternative element in $\SL(4;\mathbb{C})$ relating $R_0$ and $R_1$, i.e. $R_1 = (R_0)_{\hat P_1}$, and the analytic continuation used is
\begin{equation}\label{something2}
(t,\psi_1,\rho,x) \to (-\psi_1,-t,i\sqrt{1+\rho^2},-i x) \ .
\end{equation}
One can then show that
\begin{equation}\label{whatwhatwhat}
\hat g_0^{a.c.} = P^s g^s_{0,\lambda} h \ , \qquad
\hat g_1^{a.c.} = P^s Q^{-1} g^s_{1,\lambda} h \ , \qquad h \in \SO(1,4;\mathbb{C}) \ ,
\end{equation}
where here the analytic continuation is
\begin{equation}\label{something}
(t,\psi_1) \to i(-\psi_1,-t) \ .
\end{equation}
This analytic continuation, along with $\varkappa \to i\sdp$, maps from \eqref{backk0} to \eqref{backsplittime0} and \eqref{backk1} to \eqref{backsplittime1}, which correlates with the relations in \eqref{whatwhatwhat}.\foot{Recall that $\tilde R^s = (R^s)_Q = i(R_1)_{P^s}$, and hence $R^s = i(R_1)_{P^s Q^{-1}}$.} The first case in \eqref{whatwhatwhat} then corresponds to the third row of the table. For the fourth row we combine the second case of \eqref{whatwhatwhat} with \eqref{g0acg1} to give
\begin{equation}
\hat g_0^{a.c.} = \hat P^s g_{1,\lambda}^s h \ , \qquad h \in \SO(1,4;\mathbb{C}) \ ,
\end{equation}
where we have defined $\hat P^s = \hat P_1 P^s Q^{-1}$, an alternative element of $\SL(4;\mathbb{C})$ relating $R_0$ and $R^s$ ($R^s = i (R_0)_{\hat P^s}$), while the analytic continuation is given by
\begin{equation}
(t,\psi_1,\rho,x) \to (it,i\psi_1,i\sqrt{1+\rho^2},- i x) \ .
\end{equation}

\

We conclude this section with a brief discussion on the application of the analytic continuations to the $\eta$ deformation of the $\AdS_5\times S^5$ superstring. In \cite{Arutyunov:2015qva} the deformed semi-symmetric space supercoset action of \cite{Delduc:2013qra,Delduc:2014kha}, with $\mathfrak{psu}(2,2|4)$ $r$ matrix based on \eqref{canonical} and \eqref{pidrot}, was expanded to quadratic order in fermions, rewritten in Green-Schwarz form, and the R-R fluxes extracted.\foot{Strictly speaking one can extract the combinations $\mathcal{F}_n = e^{\Phi} F_n$ where $\Phi$ is the dilaton and $F_n$ are the R-R fluxes. Here we will loosely refer to $\mathcal{F}_n$ as the R-R fluxes.}  It was found that there does not exist a dilaton such that the background fields satisfy the bosonic type IIB supergravity equations of motion. Whether the deformed $\AdS_5 \times S^5$ superstring can itself be understood as a string theory is therefore still an open question. However, recent progress in this direction was made in \cite{Hoare:2015gda,Hoare:2015wia,Arutyunov:2015mqj}. It transpires that T-dualising the metric, $B$-field and R-R fluxes of \cite{Arutyunov:2015qva} in all the abelian shift isometries (the complete T-dual) gives a background that is a solution of the supergravity equations \cite{Hoare:2015wia,Hoare:2015gda}. The dilaton, however, contains a piece linear in the isometric directions explaining why the original background did not satisfy the supergravity equations.

We may therefore ask if applying the analytic continuations in table \ref{tab:ac} to the background of \cite{Arutyunov:2015qva} and its complete T-dual \cite{Hoare:2015wia} gives new backgrounds that describe the deformed $\frac{\PSU(2,2|4)}{\SO(1,4) \times \SO(5)}$ semi-symmetric space sigma model for extensions of the corresponding $\mathfrak{su}(2,2)$ $r$ matrices to $\mathfrak{psl}(4|4;\mathbb{C})$. For the first two rows we find that the backgrounds have the correct reality properties (i.e. the continuation of \cite{Arutyunov:2015qva} remains real, while the continuation of \cite{Hoare:2015wia} continues to have imaginary R-R flux as a result of the timelike T-duality). This suggests that in the non-split cases the would-be $r$ matrices related to these analytically continued results preserve the real form $\mathfrak{psu}(2,2|4)$. Assuming this is true, whether these $r$ matrices are precisely those constructed with \eqref{pidrot}, or some other inequivalent extensions, remains to be seen. For the final four analytic continuations the backgrounds become complex. In particular the $B$-field of the deformed $S^5$ and various components of the R-R fluxes become imaginary. This is consistent with the statement that there is no split $r$ matrix for the real form $\mathfrak{psu}(2,2|4)$. We leave the verification or otherwise of these claims as an open question.

The analytic continuations map the isometries amongst themselves, and, hence applied to the background of \cite{Hoare:2015wia}, the dilaton still has a term that is linear in the isometric coordinates. Therefore, the analytic continuations of \cite{Arutyunov:2015qva} are also not supergravity backgrounds, but they do solve the modified supergravity equations of \cite{Arutyunov:2015mqj}.

\subsection{Relation to gauged-WZW deformations}\label{ssec:gwzw}

Thus far we have discussed inhomogeneous Yang-Baxter deformations of the symmetric space sigma model. There is another interesting deformation based on the gauged WZW model, which contains the non-abelian T-dual of the symmetric space sigma model in a particular limit \cite{Sfetsos:2013wia,Hollowood:2014rla}. The two models are related by Poisson-Lie duality, a generalisation of non-abelian T-duality to the case of ``non-commutative conservation laws'' \cite{Klimcik:1995ux,Klimcik:1995jn}, and analytic continuations \cite{Vicedo:2015pna,Hoare:2015gda,Sfetsos:2015nya,Klimcik:2015gba}. As for the inhomogeneous Yang-Baxter deformations, this gauged-WZW deformation has also been generalised to the semi-symmetric space sigma model \cite{Hollowood:2014qma}. However, in contrast to the inhomogeneous Yang-Baxter deformations, there is by now a substantial amount of evidence that this model does describe a Green-Schwarz string \cite{Sfetsos:2014cea,Demulder:2015lva,Borsato:2016zcf}.

In general the geometries of the gauged-WZW deformed models have no symmetries. However, one can take a series of limits generating abelian isometries. The resulting metrics are then related (by analytic continuation) to the complete T-dual of a corresponding non-split Yang-Baxter deformation \cite{Hoare:2015gda}. Using the supergravity solutions of the gauged-WZW deformations, this connection proved to be useful in understanding how the background of the $\eta$-deformed $\AdS_5 \times S^5$ superstring, which is not a supergravity solution, is nevertheless related to a supergravity solution.

Thus far it has largely been the case that only the relation between the gauged-WZW deformation and a certain non-split Yang-Baxter deformation has been explored. As these two deformations correspond to different reality properties of the $q$-deformation parameter ($q$ is a phase factor in the former and real in the latter) of the deformed symmetry \cite{Delduc:2013fga,Delduc:2014kha,Hollowood:2014qma,Hollowood:2015dpa} this necessarily entailed some form of analytic continuation. However, for split Yang-Baxter deformations $q$ should also be a phase factor. Therefore, it is natural to ask if the deformed backgrounds of the split Yang-Baxter deformations can be recovered as real limits of the gauged-WZW deformations for symmetry groups that admit split solutions of the mcYBe.

Here we will investigate this question for the $\AdS_3 \simeq \frac{\SO(2,2)}{\SO(1,2)}$ coset. For the details of the gauged-WZW deformation we refer the reader to the literature \cite{Sfetsos:2013wia,Hollowood:2014rla,Hoare:2015gda}. For our purposes it is enough to recall that to extract the geometry of the model we fix a gauge on the group-valued field and integrate out the gauge field. The gauge symmetry in this theory acts as follows
\begin{equation}
f \to h^{-1} f h \ ,
\end{equation}
and hence for non-compact groups we have a number of inequivalent gauge-fixings. Let us consider the basis $\{T_a,S_a\}$, $T_{a} = i^{\lfloor\frac{a-1}{2}\rfloor}(\sigma_a\oplus \sigma_a)$, $S_a = i^{\lfloor\frac{a-1}{2}\rfloor}(\sigma_a\oplus-\sigma_a)$, of the algebra $\mathfrak{so}(2,2) \simeq \mathfrak{su}(1,1) \oplus \mathfrak{su}(1,1)$, where $\sigma_a$ are the Pauli matrices. The gauge algebra is generated by the diagonal, i.e. by $T_a$. We then consider the following gauge-fixed forms of $f$
\begin{equation}\label{csctct}
f = e^{\chi_1 S_a} e^{\chi_3 T_b} e^{\chi_2 T_a} \ ,
\end{equation}
for fixed $a$ and $b$. We now have the following three choices: $(i)$ $S_a$ and $T_a$ are compact and $T_b$ is non-compact, $(ii)$ all three generators are non-compact, and $(iii)$ $S_a$ and $T_a$ are non-compact, while $T_b$ is compact.\foot{By non-compact we mean $\tr(T^2) > 0$ and by compact, $\tr(T^2) <0$, where $\tr$ denotes the usual matrix trace.} The limiting procedure of \cite{Hoare:2015gda} then amounts to shifting $\chi_1 \to \chi_1 + \log \gamma$, or $\chi_1 \to \chi_1 + i \log \gamma$ if $S_a$ is compact, and taking $\gamma \to \infty$. This generates abelian isometries corresponding to shifts in $\chi_1$ and $\chi_2$. In particular, if the generator $S_a$ is compact (case $(i)$) we generate a $\U(1)^2$ isometry, and after analytic continuation we find the complete T-dual of a non-split deformation of $\AdS_3$. On the other hand, if $S_a$ is non-compact (cases $(ii)$ and $(iii)$) we generate a $\mathbb{R}^2$ isometry and expect to find the complete T-dual of a split deformation of $\AdS_3$.

Before discussing the split deformations, let us first recall the details of case $(i)$, associated to the non-split deformation \cite{Hoare:2015gda}. The metric in this case, with $\chi_1 = t$, $\chi_2 = \psi$ and $\chi_3 = \xi$, is
\begin{align} \nonumber
ds^2 & = \frac{k}{2\pi}\frac{1-\lambda^2}{1+\lambda^2}\Big[-dt^2 + \coth^2\xi (J^2 + K^2) - \frac{4\lambda^2}{(1-\lambda^2)^2}\coth^2\xi\big(\sinh^2\xi(dt-K)^2 -J^2\big)\Big] \ ,
\\ \nonumber
J & = \csc 2t \big(\sin 2\psi \tanh\xi\, d\xi-(\cos 2t - \cos 2\psi) d\psi\big) \ , \vphantom{\Big[\Big]}
\\ \label{metcase1}
K & = \csc 2t \big((\cos2t + \cos 2\psi)\tanh\xi \, d\xi - \sin2\psi \, d\psi\big) \ . \vphantom{\Big[\Big]}
\end{align}
Here $k$ is the level of the gauged WZW model, while $\lambda$ is the deformation parameter.\foot{The parameter $\lambda$ is the same as that used in \cite{Hollowood:2014qma,Hoare:2015gda}, which is related to the parameter of \cite{Sfetsos:2013wia,Hollowood:2014rla} by $\lambda^2 \to \lambda$.} We now start by shifting $t \to t + i \log \gamma$ and taking $\gamma \to \infty$. Then we perform the following analytic continuation and field redefinition
\begin{equation}
t \to i \varkappa t + \frac{i}{2} \log\big[\frac{1+\rho^2}{1-\varkappa^2\rho^2}\big] \ , \qquad \xi \to \frac12\log\big[\frac{1-i\rho}{1+i\rho}\big] \ , \qquad
\psi \to i\varkappa \psi \ ,
\end{equation}
which can be decomposed into two steps, the multiplication of all coordinates by $i$ and a real field redefinition. Finally, we analytically continue the parameters ($k$ becomes imaginary and $\lambda^2$ a phase factor)
\begin{equation}\label{imagredef}
k = \frac{2i\pi}{\varkappa} \ , \qquad \lambda^2 = \frac{i-\varkappa}{i+\varkappa} \ .
\end{equation}
As a result we end up with the metric
\begin{equation}
ds^2 = \frac{d\rho^2}{(1-\varkappa^2\rho^2)(1+\rho^2)} - \frac{1-\varkappa^2\rho^2}{1+\rho^2}dt^2 +\frac{d\psi^2}{\rho^2} \ .
\end{equation}
Discarding total derivatives in the $B$-field this is indeed the complete T-dual\foot{Note that if we keep the total derivatives in the $B$-field, after the T-duality they will give off-diagonal terms in the metric that can be removed by a coordinate transformation.} of a non-split Yang-Baxter deformation of $\AdS_3$, as can be seen from truncating (setting $x=0$ in) \eqref{backk0}.

The simplest approach to treating cases $(ii)$ and $(iii)$ is as analytic continuations of case $(i)$. The metric for case $(ii)$ is given by \eqref{metcase1} with
\begin{equation}
t \to - i \psi_1 \ , \qquad \psi \to -i \psi_2 \ .
\end{equation}
To take the limit in this case we shift $\psi_1 \to \psi_1 + \log \gamma$ and send $\gamma \to \infty$. We then use the following real field redefinition
\begin{equation}
\psi_1 \to \sdp \psi_1 + \frac{1}{2} \log\big[\frac{1-t^2}{1-\sdp^2t^2}\big] \ , \qquad \xi \to \frac12\log\big[\frac{1-t}{1+t}\big] \ , \qquad
\psi_2 \to \sdp \psi_2 \ ,
\end{equation}
and real redefinition of the parameters\foot{We refer to this as a real redefinition of the parameters as only $\lambda^2$ appears in the bosonic model. In the deformation of of the semi-symmetric space sigma model \cite{Hollowood:2014qma} $\lambda$ also appears and hence we would require that $\sdp < 1$. In the gauged-WZW deformations \cite{Sfetsos:2013wia,Hollowood:2014rla} the parameter $\lambda^2$ is usually restricted to take values in the range $[0,1]$, which is equivalent to $\sdp < 1$.}$^{,}$\foot{In \cite{Borsato:2016zcf} the parameter $\kappa$ is used instead of $\sdp$.}
\begin{equation}\label{realredef}
k = \frac{2\pi}{\sdp} \ , \qquad \lambda^2 = \frac{1-\sdp}{1+\sdp} \ .
\end{equation}
The resulting metric is
\begin{equation}
ds^2 = - \frac{dt^2}{(1-\sdp^2t^2)(1-t^2)} + \frac{1-\sdp^2t^2}{1-t^2}d\psi_1^2 + \frac{d\psi_2^2}{t^2} \ ,
\end{equation}
which is the complete T-dual of a split Yang-Baxter deformation of $\AdS_3$ as can be seen from truncating \eqref{backsplitspace0}.

Finally let us consider case $(iii)$. The metric in this case is given by \eqref{metcase1} with
\begin{equation}
t \to - i \psi \ , \qquad \psi \to - i t \ , \qquad \xi \to i \xi \ .
\end{equation}
As for case $(ii)$, to take the limit we shift $\psi \to \psi + \log \gamma$ and send $\gamma \to \infty$. We then use the following real field redefinition
\begin{equation}
\psi \to \sdp \psi + \frac12\log\big[\frac{1+\rho^2}{1+\sdp^2\rho^2}\big] \ , \qquad \xi \to \frac{i}{2}\log\big[\frac{1+i\rho}{1-i\rho}\big] \ , \qquad t \to \sdp t \ ,
\end{equation}
and  the real redefinition of the parameters \eqref{realredef}. The metric that we find is
\begin{equation}
ds^2 = \frac{d\rho^2}{(1+\sdp^2\rho^2)(1+\rho^2)} + \frac{1+\sdp^2\rho^2}{1+\rho^2}d\psi^2 - \frac{dt^2}{\rho^2} \ ,
\end{equation}
which is the complete T-dual of a split Yang-Baxter deformation of $\AdS_3$ as can be seen from truncating \eqref{backsplittime0}.

The results for cases $(i)$ and $(ii)$ can be easily truncated to the $\AdS_2 \simeq \frac{\SO(2,1)}{\SO(1,1)}$ coset by setting the field in \eqref{csctct} associated to $\chi_2$ equal to zero. The $\mathfrak{so}(2,1) \simeq \mathfrak{su}(1,1)$ algebra is generated by $S_a$, $S_{6-a-b}$ and $T_b$, with $T_b$ generating the gauge algebra. Therefore, as the gauge algebra $\mathfrak{so}(1,1)$ contains a single non-compact generator, case $(iii)$ cannot be truncated in this manner.

In summary, depending on the gauge fixing used in the gauged-WZW deformation for the coset $\AdS_3 \simeq \frac{\SO(2,2)}{\SO(1,2)}$ we find various different metrics covering different patches. Each of these patches admits a limit leading either to the complete T-dual of a non-split deformation or of a split deformation. In the former case we also need to analytically continue the coordinates and parameters, while in the latter case we do not. This is in accordance with the fact that the gauged-WZW and the split Yang-Baxter deformations both have a $q$-deformation parameter that is a phase factor, while the non-split deformation corresponds to real $q$. It is natural to expect that similar results hold for the $\AdS_5 \simeq \frac{\SO(2,4)}{\SO(1,4)}$ coset.

Finally, one may also consider these limits in the gauged-WZW deformation for $\AdS_3 \times S^3$, where we use the same redefinition of parameters \eqref{imagredef} or \eqref{realredef} for both the $\AdS_3$ and $S^3$ parts. In the first case we find the complete T-dual of the non-split Yang-Baxter deformation of $S^3$ \cite{Hoare:2015gda}. In the second case we find the complete T-dual of a non-real background corresponding to a split Yang-Baxter deformation of $S^3$ where the $r$ matrix does not preserve the real form of the isometry algebra $\mathfrak{so}(4)$. Due to the low dimension of $S^3$ the non-reality is restricted to the total derivative in the $B$-field, however in higher dimensions this is no longer the case \cite{Hoare:2014pna}.

\section{Light-cone gauge \texorpdfstring{$\mathbf{S}$}{S}-matrices}\label{sec:smat}

In this section we investigate the non-split deformation of the $\AdS_5 \times S^5$ superstring with the aim of comparing the different deformations of $\AdS_5$ discussed in section \ref{ssec:non-split}. To do this we consider various light-cone gauge fixings (which reduce to the BMN one \cite{Berenstein:2002jq,Frolov:2006cc} in the undeformed limit) and compute the resulting bosonic two-particle $S$-matrices at tree level following the undeformed computation \cite{Klose:2006zd}. This requires us to single out two isometric directions -- a timelike isometry in the deformed $\AdS_5$ and a spacelike isometry in the deformed $S^5$. One such example was investigated in \cite{Arutyunov:2013ega}, however, as the deformation does not treat the isometric directions on an equal footing it is natural to ask how choosing different directions affects the light-cone gauge $S$-matrix.

\subsection[Non-split deformation of \texorpdfstring{$S^5$}{S5}]{Non-split deformation of \texorpdfstring{$\mathbf{S^5}$}{S5}}\label{ssec:s5def}

The non-split deformations of $\AdS_5$ are studied in detail in section \ref{ssec:non-split}, with the three possible metrics and $B$-fields given in \eqref{backk0}, \eqref{backk1} and \eqref{backk2}. Therefore we start by recalling the details of the non-split deformation of $S^5$ \cite{Delduc:2013fga,Arutyunov:2013ega}. For our purposes the deformed $S^5$ background can be found by analytically continuing the three deformations of $\AdS_5$ as follows:
\begin{itemize}
\item $\rho \to i r$, $x \to w$, $t \to \varphi$ and $\psi_{1,2} \to \phi_{1,2}$ in \eqref{backk0},
\item $\rho \to i r$, $x \to i w$, $t \to - \varphi$ and $\psi_{1,2} \to  - \phi_{1,2}$ in \eqref{backk1} and \eqref{backk2},
\end{itemize}
along with reversing the overall sign of the metric and $B$-field. The resulting three backgrounds
\begingroup
\allowdisplaybreaks
\begin{align}
\nonumber
\widehat{ds}_0^2 & = \frac1{1+\varkappa^2r^2}\Big((1-r^2)d\varphi^2 + \frac{dr^2}{1-r^2}\Big) + \frac1{1+\varkappa^2r^4 w^2}\Big(r^2(1-w^2) d\phi_1^2 + \frac{r^2 dw^2}{1-w^2}\Big) + r^2 w^2 d\phi_2^2
\\ \label{backks0}
\widehat{B}_0 & = \frac{\varkappa r}{1+\varkappa^2r^2} d\varphi \wedge dr - \frac{\varkappa r^4 w}{1+\varkappa^2 r^4 w^2} d\phi_1 \wedge dw \ ,
\\ \nonumber
\\ \nonumber
\widehat{ds}_1^2 & =
\frac1{1+\varkappa^2(1-r^2)}\Big(r^2 d\phi_1^2 + \frac{dr^2}{1-r^2}\Big)
\\ \nonumber
 & \quad + \frac1{1+\varkappa^2(1-r^2)^2w^2}\Big((1-r^2)(1-w^2)d\varphi^2 + \frac{(1-r^2)dw^2}{1-w^2} \Big)+ (1-r^2)w^2 d\phi_2^2
\\ \label{backks1}
\widehat{B}_1 & = - \frac{\varkappa r}{1+\varkappa^2(1-r^2)} d\phi_1 \wedge dr - \frac{\varkappa (1-r^2)^2 w}{1+\varkappa^2(1-r^2)^2 w^2} d\varphi \wedge dw \ ,
\\ \nonumber
\\ \nonumber
\widehat{ds}_2^2 & = \frac1{1+\varkappa^2(1-r^2)}\Big(r^2 d\phi_2^2 + \frac{dr^2}{1-r^2}\Big)
\\ \nonumber & \quad + \frac1{1+\varkappa^2 (1-r^2)^2(1-w^2)}\Big((1-r^2)w^2d\phi_1^2 + \frac{(1-r^2)dw^2}{1-w^2}\Big) + (1-r^2)(1-w^2)d\varphi^2
\\ \label{backks2}
\widehat{B}_2 & = - \frac{\varkappa r}{1+\varkappa^2(1-r^2)}d\phi_2 \wedge dr - \frac{\varkappa (1-r^2)^2 w }{1+\varkappa^2(1-r^2)^2(1-w^2)}d\phi_1 \wedge dw \ .
\end{align}
\endgroup
are related by ($\varkappa$-independent) coordinate redefinitions for the full ranges of the non-isometric coordinates
\begin{equation}\label{rangenis}
r \in [0,1) \ , \qquad w \in [0,1) \ ,
\end{equation}
apart from isolated points. Explicitly these diffeomorphisms are given by
\begin{equation}\label{quotes1}
\varphi \to \phi_1 \ , \quad \phi_1 \to \varphi \ , \quad \phi_2 \to \phi_2 \ , \quad r \to \sqrt{1-r^2} \ , \quad w \to w \ ,
\end{equation}
mapping from \eqref{backks0} to \eqref{backks1} and
\begin{equation}\label{quotes2}
\varphi \to \phi_2 \ , \quad \phi_1 \to - \phi_1 \ , \quad \phi_2 \to \varphi \ , \quad r \to \sqrt{1-r^2} \ , \quad w \to \sqrt{1-w^2} \ ,
\end{equation}
mapping from \eqref{backks0} to \eqref{backks2}. This implies that these three backgrounds are completely equivalent, which is to be expected as there is only a single non-split $r$ matrix for $\mathfrak{su}(4) \simeq \mathfrak{so}(6)$, the isometry algebra of $S^5$. Nevertheless, in what follows we will light-cone gauge fix in different angles and as such it will be useful to have all three forms.

In the following analysis we will also use that the diffeomorphisms \eqref{coordtrans1} and \eqref{coordtrans2} can be analytically continued to diffeomorphisms between the backgrounds \eqref{backks0}, \eqref{backks1} and \eqref{backks2}. The transformation
\begin{equation}\begin{split}\label{coordtranss1}
r \to & \frac{\sqrt{1-(1-r^2)(1-w^2)}}{\sqrt{1+\varkappa ^2 (1-r^2) (1-w^2)}} \ ,
\\
w \to & \frac{r \sqrt{1+\varkappa ^2 (1-r^2) (1-w^2)}}{\sqrt{1+\varkappa ^2 (1-r^2)} \sqrt{1-(1-r^2)(1-w^2)}} \ ,
\end{split}\end{equation}
maps \eqref{backks0} to \eqref{backks2} up to a total derivative in $B$-field, while
\begin{equation}\begin{split}\label{coordtranss2}
r \to & \frac{\sqrt{1 - (1 - r^2)w^2 }}{\sqrt{1 + \varkappa ^2 (1- r^2)w^2 }} \ , \qquad \quad \varphi \to \phi_2 \ , \quad \phi_1 \to -\varphi \ , \quad \phi_2 \to \phi_1 \ ,
\\
w \to & \frac{r \sqrt{1 + \varkappa ^2 (1- r^2)w^2 }}{\sqrt{1 + \varkappa ^2 (1-r^2)}\sqrt{1-(1-r^2)w^2}} \ ,
\end{split}\end{equation}
maps \eqref{backks0} to \eqref{backks1}. In both cases the domain and range of these maps is not necessarily the full range of the non-isometric coordinates.

\subsection{Light-cone gauge}\label{ssec:lcgf}

Here our discussion will be restricted to a special sub-class of BMN-type light-cone gauges for which the string is only moving in one of the Cartan directions of $\AdS_5$ and one of $S^5$. To recall we need to single out two isometric directions -- a timelike isometry in the deformed $\AdS_5$ and a spacelike isometry in the deformed $S^5$. When picking the timelike coordinate there is then only a single choice for each background \eqref{backk0}, \eqref{backk1} and \eqref{backk2}. On the other hand when selecting the spacelike coordinate we could pick $\varphi$, $\phi_1$ or $\phi_2$ in \eqref{backks0}. Using the diffeomorphisms \eqref{quotes1} and \eqref{quotes2} we see that these three choices are equivalent to always choosing $\varphi$ and picking between the backgrounds \eqref{backks0}, \eqref{backks1} and \eqref{backks2}. Therefore we have nine possible light-cone gauge fixings, corresponding to the BMN-like solution
\begin{equation}\label{finalsol}
t = \varphi = \tau \ , \qquad \rho = r = 0 \ , \qquad (x = w = 0) \ ,
\end{equation}
where we can pair any of the three deformations of $\AdS_5$ \eqref{backk0}, \eqref{backk1}, \eqref{backk2} with any of the three deformations of $S^5$ \eqref{backks0}, \eqref{backks1}, \eqref{backks2}. Note that for \eqref{backk1} and \eqref{backk2} we need to take $x=0$ and for \eqref{backks1} and \eqref{backks2} we need to take $w=0$.

Using the diffeomorphism \eqref{coordtrans1} we can see that the expansions around \eqref{finalsol} for the backgrounds \eqref{backk0} and \eqref{backk2} will be related by a field redefinition. Similarly, using \eqref{coordtranss1}, the expansions around \eqref{finalsol} for the backgrounds \eqref{backks0} and \eqref{backks2} will also be related by a field redefinition. As such the corresponding light-cone gauges will be equivalent and we are left with four possible light-cone gauges. To summarise, these are given by expanding around \eqref{finalsol} in the following backgrounds:
\begin{itemize}
\item Case $(i)$: \eqref{backk0} and \eqref{backks0}.
\\ The isometries of the solution \eqref{finalsol} do not appear in the non-trivial part of either the deformed $\AdS_5$ or $S^5$ $B$-fields. This is the case that was studied in \cite{Arutyunov:2013ega}.
\item Case $(ii)$: \eqref{backk0} and \eqref{backks1}
\\ The isometries of the solution \eqref{finalsol} do not appear in the deformed $\AdS_5$ $B$-field, but do appear in the deformed $S^5$ $B$-field.
\item Case $(iii)$: \eqref{backk1} and \eqref{backks0}
\\ The isometries of the solution \eqref{finalsol} do appear in the deformed $\AdS_5$ $B$-field, but do not appear in the deformed $S^5$ $B$-field.
\item Case $(iv)$: \eqref{backk1} and \eqref{backks1}
\\ The isometries of the solution \eqref{finalsol} appear in both the deformed $\AdS_5$ and $S^5$ $B$-fields.
\end{itemize}
Cases $(i)$ and $(ii)$ are light-cone gauges in region I of the deformed $\AdS_5$ space, while cases $(iii)$ and $(iv)$ correspond to region II. As they do not contribute to the light-cone gauge $S$-matrices we ignore the total derivative terms in the $B$-fields. However, for technical reasons, it is useful to add the total derivative term $-\frac{\varkappa x}{1-\varkappa^2x^2}dt\wedge dx$ to \eqref{backk1} and $\frac{\varkappa w}{1+\varkappa^2 w^2}d\varphi \wedge dw$ to \eqref{backks1}.

To proceed we substitute the various backgrounds into the light-cone gauge-fixed Lagrangian \cite{Arutyunov:2014jfa}. In particular, writing
\begin{equation}
x^+ = (1-a) t + a \varphi \ , \qquad x^- = \varphi - t \ ,
\end{equation}
where $a \in [0,1]$ is a free parameter, we T-dualise in $x^-$ and fix\foot{In the literature this is understood to be equivalent to the usual light-cone gauge-fixing, $x^+ = \tau$, $p_- = 1$ \cite{Frolov:2006cc,Arutyunov:2009ga}.}
\begin{equation}
x^+ = \tau \ , \qquad \hat x^- = \sigma \ ,
\end{equation}
where $\hat x^-$ is the T-dual coordinate of $x^-$ \cite{Kruczenski:2004cn,Zarembo:2009au,Arutyunov:2014jfa}.
We then map to cartesian coordinates in the neighbourhood of the expansion points and expand to quartic order in the fields. Explicitly, for the $\AdS_5$ part, if the background we are considering involves \eqref{backk0} we set
\begin{equation}
Z^{3\dot 3} = \frac{\rho\sqrt{1-x^2}}{\sqrt{2}} e^{-i\psi_1} \ , \qquad Z^{4\dot 4} = (Z^{3\dot 3})^* \ , \qquad
Z^{3\dot 4} = \frac{\rho x}{\sqrt{2}} e^{-i\psi_2} \ , \qquad Z^{4\dot 3} = - (Z^{3\dot 4})^* \ ,
\end{equation}
while if it involves \eqref{backk1} we set
\begin{equation}\label{coc2}
Z^{3\dot 3} = \frac{x}{\sqrt{2}\sqrt{1+\varkappa^2}} e^{-i\psi_1} \ ,  \qquad Z^{4\dot 4} = (Z^{3\dot 3})^* \ , \qquad
Z^{3\dot 4} = \frac{\rho}{\sqrt{2}} e^{-i\psi_2} \ ,  \qquad Z^{4\dot 3} = - (Z^{3\dot 4})^* \ .
\end{equation}
Similarly, for the $S^5$ part, if the background involves \eqref{backks0} we set
\begin{equation}
Y^{1\dot 1} = \frac{r\sqrt{1-w^2}}{\sqrt{2}} e^{i\phi_1} \ , \qquad Y^{2\dot 2} = (Y^{1\dot 1})^* \ , \qquad
Y^{1\dot 2} = \frac{r w}{\sqrt{2}} e^{-i\phi_2} \ , \qquad Y^{2\dot 1} = - (Y^{1\dot 2})^* \ ,
\end{equation}
and if it involves \eqref{backks1} we set
\begin{equation}\label{coc4}
Y^{1\dot 1} = \frac{w}{\sqrt{2}\sqrt{1+\varkappa^2}} e^{i\phi_1} \ ,  \qquad Y^{2\dot 2} = (Y^{1\dot 1})^* \ , \qquad
Y^{1\dot 2} = \frac{r}{\sqrt{2}} e^{-i\phi_2} \ ,  \qquad Y^{2\dot 1} = - (Y^{1\dot 2})^* \ .
\end{equation}
The reason for these differences is that for backgrounds \eqref{backk0} and \eqref{backks0} close to the solution \eqref{finalsol} $x$ and $w$ play the role of angles, while for \eqref{backk1} and \eqref{backks1} they play the role of radial coordinates. The additional rescaling of $x$ and $w$ by $\sqrt{1+\varkappa^2}$ in \eqref{coc2} and \eqref{coc4} is included in order that the quadratic terms of all four light-cone gauge Lagrangians are the same.
From the resulting Lagrangian we can then compute the bosonic two-particle tree-level $S$-matrix.

In the following we use the standard notation, writing $Y^{a\dot a}$ and $Z^{\alpha \dot \alpha}$ ($a = 1,2$, $\dot a = \dot 1,\dot 2$, $\alpha=3,4$ and $\dot \alpha = \dot3,\dot4$), which satisfy the following reality conditions
\begin{equation}
(Y^{a\dot a})^* = \epsilon_{ab}\epsilon_{\dot a \dot b} Y^{b\dot b} \ , \qquad (Z^{\alpha\dot \alpha})^* = \epsilon_{\alpha \beta} \epsilon_{\dot \alpha \dot \beta} Z^{\beta\dot \beta} \ ,
\end{equation}
where $\epsilon$ is the two-index antisymmetric tensor ($\epsilon_{12} = \epsilon_{\dot 1\dot 2} = \epsilon_{34} = \epsilon_{\dot 3 \dot 4} = 1$).
In the undeformed theory the indices $a$, $\dot a$, $\alpha$ and $\dot \alpha$ are fundamental indices of the bosonic light-cone gauge symmetry algebra $\mathfrak{su}(2)^4$. Correspondingly we raise and lower these indices with $\epsilon$. This symmetry is broken to its Cartan subgroup by the deformation, however it is still useful to use the associated structure. The remaining $\U(1)^4$ symmetry acts as
\begin{equation}\begin{split}\label{u14sym}
Y^{1\dot 1} \to e^{i\varepsilon_1} Y^{1\dot 1} \ , \qquad
Y^{1\dot 2} \to e^{i\varepsilon_2} Y^{1\dot 2} \ , & \qquad
Z^{3\dot 3} \to e^{i\varepsilon_3} Z^{3\dot 3} \ , \qquad
Z^{3\dot 4} \to e^{i\varepsilon_4} Z^{3\dot 4} \ .
\end{split}\end{equation}

We now present the expressions for the quadratic and quartic Lagrangians. For all four cases listed above the quadratic Lagrangian is the same and is given by\foot{We use the notation $Y^2 = Y_{a\dot a}Y^{a \dot a}$, $Y \cdot \dot Y = Y_{a\dot a} \dot Y^{a\dot a}$ and so on.}
\begin{equation}
\mathcal{L}_2 = \frac{1}{2}\big[
  \dot Y^2 - Y'^2 - (1+\varkappa^2) Y^2 +
  \dot Z^2 - Z'^2 - (1+\varkappa^2) Z^2
\big] \ .
\end{equation}
As usual $\dot{}$ and ${}'$ denote derivatives with respect to $\tau$ and $\sigma$ respectively. The free dispersion relation is therefore given by\foot{In our conventions the worldsheet momentum naturally is normalised such that the near-BMN dispersion relation is relativistic. Compared to \cite{Arutyunov:2013ega} we therefore have that $\p_{here} = \sqrt{1+\varkappa^2} \, \p_{there}$.}
\begin{equation}\label{fdp}
\e^2 = \p^2 + (1+\varkappa^2) \ ,
\end{equation}
for all eight bosonic degrees of freedom. Furthermore, the light-cone gauge Lagrangians of all four cases have a common $\SU(2)^4$ invariant part, while the terms breaking this symmetry (which vanish on setting $\varkappa = 0$) are different. The common $\SU(2)^4$ invariant part is given by\foot{Here we have performed an additional field redefinition that has been chosen to remove all $Y^4$ and $Z^4$ terms (not containing derivatives). This field redefinition respects the $\U(1)^4$ symmetry \eqref{u14sym} of the deformed theory. Further, if we set $\varkappa = 0$ it preserves the $\SU(2)^4$ symmetry of the undeformed model.}
\begin{align}\nonumber
\mathcal{L}_{\SU(2)^4} & = \frac12\big[Y^2 Y'^2 + \varkappa^2 Y^2 \dot Y^2  - Z^2 Z'^2 - \varkappa^2 Z^2 \dot Z^2\big] + \frac{1}{4}(1+\varkappa^2) \big[Y^2 (\dot Z^2 + Z'^2) - Z^2 (\dot Y^2 + Y'^2) \big]
\\\nonumber
& + \frac18 (1-2a) \big[
\big(Y_+^2 + Z_+^2\big)\big(Y_-^2 + Z_-^2\big) -
 (1+\varkappa^2) \big(Y_+ \cdot Y_- + Z_+ \cdot Z_- \big)\big(Y^2 + Z^2)
\\\label{lag4sym} & \hspace{145pt} \vphantom{\frac18}
 - 2(1+\varkappa^2)\big(Y\cdot Y_+ + Z \cdot Z_+\big)\big(Y\cdot Y_- + Z \cdot Z_-\big)\big] \ ,
\end{align}
where we have defined $Y_\pm = \dot Y \pm Y'$ and $Z_\pm = \dot Z \pm Z'$.
After integrating by parts and dropping total derivatives the four symmetry-breaking terms are given by
\begingroup
\allowdisplaybreaks
\begin{align}
\label{lag4breaki}
\mathcal{L}_{(i)} & =
    4 i \varkappa \, Y^{1\dot 2}Y^{2\dot 1} (\dot Y^{1\dot 1} Y'^{2\dot 2} - Y'^{1\dot 1}\dot Y^{2\dot 2})
  + 4 i \varkappa \, Z^{3\dot 4}Z^{4\dot 3} (\dot Z^{3\dot 3} Z'^{4\dot 4} -  Z'^{3\dot 3}\dot Z^{4\dot 4}) \ ,
\\ & \nonumber
\\ \label{lag4breakii}
\begin{split}
\mathcal{L}_{(ii)} & =
  -   4 \varkappa (1+\varkappa^2) Y^{1\dot 1}Y^{2\dot 2}(Y^{1\dot 2} Y^{2\dot 1})'
    - 2\varkappa^2(Y^{1\dot 1}Y^{2\dot 2})\dot{}(Y^{1\dot 2}Y^{2\dot 1})\dot{} + 2\varkappa^2(Y^{1\dot 1}Y^{2\dot 2})'(Y^{1\dot 2}Y^{2\dot 1})'
\\ & + 3\varkappa^2 Y^{1\dot 1} Y^{2\dot 2}(\dot Y^{1\dot 2} \dot Y^{2\dot 1} - Y'^{1\dot 2}Y'^{2\dot 1})
    - 3\varkappa^2 Y^{1\dot 2} Y^{2\dot 1}(\dot Y^{1\dot 1} \dot Y^{2\dot 2} - Y'^{1\dot 1}Y'^{2\dot 2})
\\ & + 4 i \varkappa \, Z^{3\dot 4}Z^{4\dot 3} (\dot Z^{3\dot 3} Z'^{4\dot 4} -  Z'^{3\dot 3}\dot Z^{4\dot 4}) \ ,
\end{split}
\\ & \nonumber
\\ \label{lag4breakiii}
\begin{split}
\mathcal{L}_{(iii)} & =
    4 i \varkappa \, Y^{1\dot 2}Y^{2\dot 1} (\dot Y^{1\dot 1} Y'^{2\dot 2} - Y'^{1\dot 1}\dot Y^{2\dot 2})
\\ & - 4 \varkappa (1+\varkappa^2) Z^{3\dot 3}Z^{4\dot 4}(Z^{3\dot 4} Z^{4\dot 3})'
    + 2\varkappa^2(Z^{3\dot 3}Z^{4\dot 4})\dot{}(Z^{3\dot 4}Z^{4\dot 3})\dot{} - 2\varkappa^2(Z^{3\dot 3}Z^{4\dot 4})'(Z^{3\dot 4}Z^{4\dot 3})'
\\ & - 3\varkappa^2 Z^{3\dot 3} Z^{4\dot 4}(\dot Z^{3\dot 4} \dot Z^{4\dot 3} - Z'^{3\dot 4}Z'^{4\dot 3})
    + 3\varkappa^2 Z^{3\dot 4} Z^{4\dot 3}(\dot Z^{3\dot 3} \dot Z^{4\dot 4} - Z'^{3\dot 3}Z'^{4\dot 4}) \ ,
\end{split}
\\ & \nonumber
\\ \label{lag4breakiv}
\begin{split}
\mathcal{L}_{(iv)} & =
    -4 \varkappa (1+\varkappa^2) Y^{1\dot 1}Y^{2\dot 2}(Y^{1\dot 2} Y^{2\dot 1})'
    - 2\varkappa^2(Y^{1\dot 1}Y^{2\dot 2})\dot{}(Y^{1\dot 2}Y^{2\dot 1})\dot{} + 2\varkappa^2(Y^{1\dot 1}Y^{2\dot 2})'(Y^{1\dot 2}Y^{2\dot 1})'
\\ & + 3\varkappa^2 Y^{1\dot 1} Y^{2\dot 2}(\dot Y^{1\dot 2} \dot Y^{2\dot 1} - Y'^{1\dot 2}Y'^{2\dot 1})
    - 3\varkappa^2 Y^{1\dot 2} Y^{2\dot 1}(\dot Y^{1\dot 1} \dot Y^{2\dot 2} - Y'^{1\dot 1}Y'^{2\dot 2})
\\ & - 4 \varkappa (1+\varkappa^2) Z^{3\dot 3}Z^{4\dot 4}(Z^{3\dot 4} Z^{4\dot 3})'
    + 2\varkappa^2(Z^{3\dot 3}Z^{4\dot 4})\dot{}(Z^{3\dot 4}Z^{4\dot 3})\dot{} - 2\varkappa^2(Z^{3\dot 3}Z^{4\dot 4})'(Z^{3\dot 4}Z^{4\dot 3})'
\\ & - 3\varkappa^2 Z^{3\dot 3} Z^{4\dot 4}(\dot Z^{3\dot 4} \dot Z^{4\dot 3} - Z'^{3\dot 4}Z'^{4\dot 3})
    + 3\varkappa^2 Z^{3\dot 4} Z^{4\dot 3}(\dot Z^{3\dot 3} \dot Z^{4\dot 4} - Z'^{3\dot 3}Z'^{4\dot 4}) \ .
\end{split}
\end{align}
\endgroup
For example, the light-cone gauge action for case $(i)$ to quartic order in fields is given by
\begin{equation}
\mathcal{S}_{(i)} = T \int d\tau d\sigma \, \big[\mathcal{L}_2 + \mathcal{L}_{\SU(2)^4} + \mathcal{L}_{(i)} + \ldots \big] \ ,
\end{equation}
where $T$ is the effective string tension. Similar expressions hold for the remaining three cases. Rescaling the fields by $\frac{1}{\sqrt{T}}$ we then compute the bosonic two-particle tree-level $S$-matrix, i.e. the leading term in $\mathbb{T}$, which is related to the $S$-matrix as follows
\begin{equation}
\mathbb{S} = \mathbb{I} + \frac{i}{T} \mathbb{T} \ .
\end{equation}

\subsection[\texorpdfstring{$S$}{S}-matrices]{\texorpdfstring{$\mathbf{S}$}{S}-matrices}\label{ssec:smatres}

To present the results we recall that the full theory also contains the fermionic degrees of freedom $\zeta^{a\dot \alpha}$ and $\chi^{\alpha \dot a}$. Together with $Y^{a\dot a}$ and $Z^{\alpha\dot \alpha}$ these can be grouped into a single object $\Phi^{A\dot A}$ where $A =(a,\alpha)$ and $\dot A = (\dot a, \dot \alpha)$. It was shown in \cite{Arutyunov:2013ega} that in case $(i)$ the {\em bosonic} tree-level $S$-matrix can be written as
\begin{equation}\begin{split}\label{shorthand}
\mathbb{T} |\Phi_{A\dot A}(\p) \Phi_{B\dot B}(\p')\rangle
 & =
T_{A\dot A,B\dot B}^{C\dot C,D\dot D}  |\Phi_{C\dot C}(\p) \Phi_{D\dot D}(\p')\rangle
\\
T_{A\dot A,B\dot B}^{C\dot C,D\dot D}  & =
(-1)^{[\dot A]([B] + [D])} \mathcal{T}_{AB}^{CD}\delta_{\dot A}^{\dot C}\delta_{\dot B}^{\dot D}
+ (-1)^{([\dot A]+[\dot C])[B]} \delta_A^C \delta_B^D \mathcal{T}_{\dot A\dot B}^{\dot C \dot D} \ ,
\end{split}\end{equation}
where $[a] = [\dot a] = 0$ and $[\alpha] = [\dot \alpha] = 1$. In the undeformed theory this form is a consequence of the factorisation of the full $S$-matrix into two $\mathfrak{psu}(2|2)\ltimes \mathbb{R}^3$ invariant parts. However, in the deformed model this structure appears to no longer hold for amplitudes involving fermionic external states already at tree-level \cite{Arutyunov:2015qva}, an issue we will comment upon later. Therefore, here we just use \eqref{shorthand} as a shorthand for presenting the bosonic results. The relevant entries of $\mathcal{T}_{AB}^{CD}$ are
\begin{equation}\begin{split}
\mathcal{T}_{ab}^{cd} &  = A \delta_a^c\delta_b^d + B \delta_a^d\delta_b^c + W \epsilon_{ab}\delta_a^d\delta_b^c \ ,
\\
\mathcal{T}_{\alpha\beta}^{\gamma\delta} &  = D \delta_\alpha^\gamma\delta_\beta^\delta + E \delta_\alpha^\delta\delta_\beta^\gamma + W \epsilon_{\alpha\beta}\delta_\alpha^\delta\delta_\beta^\gamma \ ,
\\
\mathcal{T}_{a\beta}^{c\delta} & = G \delta_a^c \delta_\beta^\delta \ , \qquad
\mathcal{T}_{\alpha b}^{\gamma d} = L \delta_\alpha^\gamma \delta_c^d \ ,
\end{split}\end{equation}
where at tree-level the parametrising functions are given by \cite{Arutyunov:2013ega}
\begin{equation}\begin{split}
A(\p,\p') & = \frac{1-2a}{4} (\e' \p - \e \p') +\frac{(\p-\p')^2 + \varkappa^2 (\e-\e')^2}{4(\e'\p-\e\p')}    \ ,
\\
D(\p,\p') & = \frac{1-2a}{4} (\e' \p - \e \p') -\frac{(\p-\p')^2 + \varkappa^2 (\e-\e')^2}{4(\e'\p-\e\p')}    \ ,
\\
G(\p,\p') & = \frac{1-2a}{4}(\e'\p-\e\p') -\frac{(1+\varkappa^2)(\e^2 -\e'^2 +\p^2 -\p'^2)}{8(\e'\p-\e\p')}  \ ,
\\
L(\p,\p') & = \frac{1-2a}{4}(\e'\p-\e\p') +\frac{(1+\varkappa^2)(\e^2 -\e'^2 +\p^2 -\p'^2)}{8(\e'\p-\e\p')}  \ ,
\\
B(\p,\p') & = - E(\p,\p') = \frac{\p\p'+\varkappa^2 \e\e'}{\e'\p-\e\p'}  \ , \qquad
W(\p,\p') =  i\varkappa  \ .
\end{split}\end{equation}
These expressions match \cite{Arutyunov:2013ega} a near-BMN-type expansion of the $\mathcal{U}_q(\mathfrak{psu}(2|2)) \ltimes \mathbb{R}^3$ invariant $S$-matrix of \cite{Beisert:2008tw,Beisert:2010kk,Hoare:2011wr} with the following map between parameters \cite{Arutyunov:2013ega,Delduc:2014kha,Arutynov:2014ota,Hoare:2014oua}
\begin{equation}
q = \exp \big[-\frac{\varkappa}{T} \big] \ , \qquad \xi = i \varkappa \ ,
\end{equation}
where the parameter $\xi$ was introduced in \cite{Beisert:2011wq}. By construction this $S$-matrix is a deformation of the $\mathfrak{psu}(2|2) \ltimes \mathbb{R}^3$ invariant $S$-matrix \cite{Beisert:2005tm,Beisert:2006ez,Arutyunov:2006yd,Dorey:2007xn} that underlies the scattering of excitations in the BMN light-cone gauge $\AdS_5 \times S^5$ superstring theory and therefore is a natural candidate with which to compare.

The bosonic tree-level light-cone gauge $S$-matrix in the four cases of interest can then be written as
\begin{equation}
T_{(*)}{}_{A\dot A,B\dot B}^{C\dot C,D\dot D}
 =  U_{(*)}{}_{A\dot A}^{\underline{A}\dot{\underline{A}}}
\,  U_{(*)}{}_{B\dot B}^{\underline B \dot{\underline{B}}}
\,  T_{\underline{A}\dot{\underline{A}},\underline{B}\dot{\underline{ B}}}^{\underline{C}\dot{\underline{C}},\underline{D}\dot{\underline{ D}}}
\,  U^{-1}_{(*)}{}_{\underline{C}\dot{\underline C}}^{C\dot C}
\,  U^{-1}_{(*)}{}_{\underline{D}\dot{\underline{D}}}^{D\dot D} \ ,
\end{equation}
where the matrices $U_{(*)}$ are {\em diagonal} momentum-dependent one-particle changes of basis whose
non-trivial entries are
\begin{align}
U_{(ii)}{}_{1\dot 1}^{1\dot 1} = U_{(iv)}{}_{1\dot 1}^{1\dot 1} & = \sqrt{1+\varkappa^2} \frac{\p-i\varkappa}{\p + i \varkappa \e} \ , &
U_{(ii)}{}_{2\dot 2}^{2\dot 2} = U_{(iv)}{}_{2\dot 2}^{2\dot 2} & = \sqrt{1+\varkappa^2} \frac{\p-i\varkappa}{\p - i \varkappa \e} \ , &
\end{align}
\begin{align}
U_{(iii)}{}_{3\dot 3}^{3\dot 3} = U_{(iv)}{}_{3\dot 3}^{3\dot 3} & = \sqrt{1+\varkappa^2} \frac{\p+i\varkappa}{\p - i \varkappa \e} \ , &
U_{(iii)}{}_{4\dot 4}^{4\dot 4} = U_{(iv)}{}_{4\dot 4}^{4\dot 4} & = \sqrt{1+\varkappa^2} \frac{\p+i\varkappa}{\p + i \varkappa \e} \ . &
\end{align}
The remaining diagonal entries are all equal to one. Using the dispersion relation \eqref{fdp} one can check that these changes of basis are unitary, i.e. $U_{(*)}^\dagger U^{\vphantom{\dagger}}_{(*)} = 1$ for real momentum and energy.

Therefore, expanding around the various BMN-type solutions of the deformed model we find different local geometries and correspondingly different light-cone gauge $S$-matrices. In particular, for the final three cases, the structure \eqref{shorthand} no longer holds. However, they are related by momentum-dependent one-particle changes of basis. Correspondingly, they possess the same underlying symmetry, with the differences understood as coming from different representations of that symmetry. For the same reason these tree-level $S$-matrices all satisfy the classical Yang-Baxter equation.

It would be interesting to investigate whether these changes of basis are related to the equivalence discussed in section \ref{ssec:sugra} between the deformed models based on $(R_X,g)$ and $(R,Xg)$. Naively $g \to Xg$ amounts to a local field redefinition, however, understanding the interplay with the fixing of the light-cone gauge may yield a connection to the one-particle changes of basis above.

Finally, let us note that the need for changes of basis in this theory has appeared before in two different contexts \cite{Engelund:2014pla,Arutyunov:2015qva}. In \cite{Engelund:2014pla} the authors consider the object $\mathcal{T}_{AB}^{CD}$ and assume the remaining tree-level amplitudes are those that follow from symmetry \cite{Beisert:2008tw,Beisert:2010kk,Hoare:2011wr}. They then construct the logarithms of the one-loop $S$-matrix via unitarity cuts \cite{Bianchi:2013nra,Engelund:2013fja}. The result is related to that following from symmetry by a momentum-dependent one-particle change of basis. This change of basis has a different structure to those above as it is perturbative in the coupling, that is the identity plus an $\mathcal{O}(T^{-1})$ term. On the other hand, in \cite{Arutyunov:2015qva} the authors compute the tree-level components of $T_{(i)}{}_{A\dot A,B\dot B}^{C\dot C,D\dot D}$ involving two bosons and two fermions. This is the case for which the bosonic amplitudes are consistent with the factorisation structure \eqref{shorthand}. The amplitudes involving fermions break this structure, but, as we have seen, factorisation does not necessarily need to be manifest. More curious is that the classical Yang-Baxter equation is not satisfied. To restore both properties the authors construct a change of basis of two-particle states. The meaning of this change of basis and its relation to the believed integrability of the model is an open question.

\section{Concluding remarks}\label{sec:conclusions}

In this article we have studied inequivalent Yang-Baxter deformations of $\AdS_5$ of non-split and split type. In particular, we have explored the differences between the resulting backgrounds, in which ways they are related and to what extent they admit contraction limits. In the second half of the paper we computed the bosonic two-particle tree-level $S$-matrices based on four inequivalent BMN-type light-cone gauges. The resulting $S$-matrices, while different, are related by momentum-dependent one-particle changes of basis. While the results in this paper do not directly resolve the puzzles of \cite{Arutyunov:2015qva} they do demonstrate that the choice of $r$ matrix affects certain aspects of the physics. This raises a number of interesting questions.

In section \ref{ssec:sugra} we demonstrated that the analytic continuations between the various inequivalent $r$ matrices are equivalent to analytic continuations between the backgrounds, where the latter is independent of the deformation parameter. We then conjectured that these analytic continuations may also be applied to the R-R fluxes of \cite{Arutyunov:2015qva,Hoare:2015wia} and correspond to certain extensions of the $\mathfrak{su}(2,2)$ $r$ matrices to $\mathfrak{psu}(2,2|4)$. It would be interesting to make this statement precise and investigate whether the method has further applicability.

In section \ref{ssec:gwzw}, considering the truncation to $\AdS_3$, we generalised the construction of \cite{Hoare:2015gda} to recover the T-duals of split Yang-Baxter deformations as real limits of the gauged-WZW deformation \cite{Sfetsos:2013wia,Hollowood:2014rla}. These limits are taken at the level of the geometry and it remains to be seen whether the models defined in terms of group-valued fields can be related in this manner.

It is by now a common feature of $S$-matrix computations in these models that changes of basis \cite{Engelund:2014pla,Arutyunov:2015qva} are required to match the result following from symmetries \cite{Beisert:2008tw,Hoare:2011wr}. In \cite{Engelund:2014pla} and in the results of section \ref{ssec:smatres} these are one-particle (momentum-dependent) transformations and hence integrability, i.e. the classical Yang-Baxter equation, is unaffected. It remains to be understood how the two-particle transformation needed when also scattering fermions \cite{Arutyunov:2015qva} is consistent with the classical integrability of theory. It may also be instructive to explore the remaining BMN-type light-cone gauges outlined in the appendix. These are structurally different to those considered in section \ref{ssec:lcgf} and therefore may correspond to new light-cone gauge symmetry algebras.

To conclude, in this article we have explored the effect of inequivalent $\mathfrak{su}(2,2)$ $r$ matrices. For the superalgebra $\mathfrak{psu}(2,2|4)$ there are additional, potentially inequivalent, choices that coincide on restricting to $\mathfrak{su}(2,2)$, for example based on the various Dynkin diagrams of the supersymmetry algebra. It still remains important to understand if working with these alternatives could resolve the puzzles of \cite{Arutyunov:2015qva}.

\section*{Acknowledgments}

We would like to thank S. Frolov, T. McLoughlin and A.A. Tseytlin for related discussions.
The work of B.H. is partially supported by grant no. 615203 from the European Research Council under the FP7. S.T. is supported by L.T. The work of S.T. is supported by the Einstein Foundation Berlin in the framework of the research project "Gravitation and High Energy Physics" and acknowledges further support from the People Programme (Marie Curie Actions) of the European Union's Seventh Framework Programme FP7/2007-2013/ under REA Grant Agreement No 317089.

\appendix

\section{BMN-like solutions of the deformed \texorpdfstring{$\mathbf{AdS_5 \times S^5}$}{AdS5 x S5} models}\label{ssec:bmn}
\def\theequation{A.\arabic{equation}}
\setcounter{equation}{0}

In $\AdS_5 \times S^5$ embedding coordinates
\begin{equation}\label{embeddc}
-Z_0^2 - Z_5^2 + \sum_{i=1}^4 Z_i^2 = - 1 \ , \qquad \sum_{i=1}^6 Y_i^2 = 1 \ ,
\end{equation}
the BMN solution of the undeformed string sigma-model is a point-like string moving on a great circle of $S^5$. In conformal gauge it is given by
\begin{equation}\label{bmnundef}
Z_0 + i Z_5 = e^{i \tau} \ , \qquad Y_1 + i Y_2 = e^{i \tau} \ ,
\end{equation}
where $\tau$ is the worldsheet time coordinate. The solution can be freely rotated by the global $\SO(2,4) \times \SO(6)$ isometry of $\AdS_5 \times S^5$. However, this symmetry is broken by the deformation and hence different orientations will be modified differently. In this appendix we will briefly explore which solutions survive the deformation in a certain manner to be prescribed.

\

We start by considering the deformation of $S^5$, the metric and $B$-field of which \cite{Arutyunov:2013ega} are given in \eqref{backks0} (or equivalently \eqref{backks1} and \eqref{backks2}). One convenient approach to investigating BMN string solutions is to introduce the following form of this background
\begin{align}
\nonumber
ds^2 = & f\, \mathcal{Y}_1^2 d\varphi^2 + g \, \mathcal{Y}_2^2 d\phi_1^2 + \mathcal{Y}_3^2 d\phi_2^2 + \frac{f - g \, \mathcal{Y}_1^2}{1-\mathcal{Y}_1^2} d\mathcal{Y}_1^2 + g \, d \mathcal{Y}_2^2 + g \, d \mathcal{Y}_3^2 \ ,
\\ \nonumber
B = & \frac{\varkappa}{2} \big( f \, d\varphi \wedge d(\mathcal{Y}_1^2) + g \, \mathcal{Y}_3^2 \, d\phi_1 \wedge d(\mathcal{Y}_2^2) - g \, \mathcal{Y}_2^2 \, d\phi_1 \wedge d(\mathcal{Y}_3^2)\big) \ ,
\\\label{backemb}
f = & \frac{1}{1+\varkappa^2(1-\mathcal{Y}_1^2)} \ , \qquad g = \frac{1}{1 + \varkappa^2 (1-\mathcal{Y}_1^2)\mathcal{Y}_3^2} \ ,
\end{align}
where $\mathcal{Y}_{1,2,3}$ are subject to the constraint\foot{\label{footmaps}Up to signs, to recover \eqref{backks0} we can set $\mathcal{Y}_1 = \sqrt{1-r^2}$, $\mathcal{Y}_2 = r \sqrt{1-w^2}$ and $\mathcal{Y}_3 = r w$. We can also recover \eqref{backks1} setting $\mathcal{Y}_1 = r$, $\mathcal{Y}_2 = \sqrt{1-r^2} \sqrt{1-w^2}$, $\mathcal{Y}_3 = \sqrt{1-r^2} w$ and interchanging $\varphi$ and $\phi_1$. Similarly, to recover \eqref{backks2} we set $\mathcal{Y}_1 = r$, $\mathcal{Y}_2 = \sqrt{1-r^2} w$, $\mathcal{Y}_3 = \sqrt{1-r^2} \sqrt{1-w^2}$ and interchange $\varphi$ and $\phi_2$.}
\begin{equation}\label{constraint}
\mathcal{Y}_1^2+\mathcal{Y}_2^2+\mathcal{Y}_3^2 = 1 \ .
\end{equation}
Setting $\varkappa = 0$ the $B$-field vanishes, while the metric reduces to
\begin{equation}
ds^2 = \mathcal{Y}_1^2 d\varphi^2 + \mathcal{Y}_2^2 d\phi_1^2 + \mathcal{Y}_3^2 d\phi_2^2 + d \mathcal{Y}_1^2 + d\mathcal{Y}_2^2 + d\mathcal{Y}_3^2 \ ,
\end{equation}
and we can easily relate back to the embedding coordinates \eqref{embeddc}
\begin{equation}
Y_1 + i Y_2 = \mathcal{Y}_1 \, e^{i \varphi} \ , \qquad
Y_3 + i Y_4 = \mathcal{Y}_2 \, e^{i \phi_1} \ , \qquad
Y_5 + i Y_6 = \mathcal{Y}_3 \, e^{i \phi_2} \ . \qquad
\end{equation}

The particular solutions we will look for are those that in the undeformed limit become a standard BMN string solution \eqref{bmnundef}, and remain point-like in the deformed theory. We further require that the string is solely moving in the Cartan directions (i.e. those directions that remain isometries) in the deformed theory. Taking these requirements into account we assume the ansatz
\begin{equation}\label{ansatz}
\varphi = \omega \tau \ , \quad \phi_1 = \omega \tau \ , \quad \phi_2 = \omega \tau \ , \qquad \mathcal{Y}_{1,2,3} = \text{constant} \ .
\end{equation}
In conformal gauge the contribution from the deformed sphere part of the geometry to the Virasoro constraints is given by\foot{We use light-cone coordinates on the worldsheet, $\partial_\pm = \partial_\tau \pm \partial_\sigma$.}
\begin{equation}\label{vir}
\nu_{_S} = G^S_{\mu\nu} \partial_\pm X^\mu \partial_\pm X^\nu = \big(\frac{\mathcal{Y}_1^2}{1+\varkappa^2(1-\mathcal{Y}_1^2)}+\frac{\mathcal{Y}_2^2}{1+\varkappa^2(1-\mathcal{Y}_1^2)\mathcal{Y}_3^2} + \mathcal{Y}_3^2\big) \, \omega^2 \ ,
\end{equation}
where $G^S_{\mu\nu}$ is the metric of the deformed $S^5$. Substituting the ansatz \eqref{ansatz} into the equations of motion and fixing $\omega$ so that $\nu_{_S} = 1$, we find the following solutions\foot{Recall that to impose the constraint \eqref{constraint} one should add a Lagrange multiplier to the action in the usual manner.}
\begingroup
\allowdisplaybreaks
\begin{align}
\label{sol1}
& \mathcal{Y}_1 = 1 \ , \qquad \mathcal{Y}_{2,3} = 0 \ , \qquad \omega = 1 \ ,
\\
\label{sol2}
& \mathcal{Y}_2 = 1 \ , \qquad \mathcal{Y}_{1,3} = 0 \ , \qquad \omega = 1 \ ,
\\
\label{sol3}
& \mathcal{Y}_3 = 1 \ , \qquad \mathcal{Y}_{1,2} = 0 \ , \qquad \omega = 1 \ ,
\\
\label{sol4}
& \mathcal{Y}_1 = \sqrt[4]{1+\varkappa^2} \mathcal{Y}_2 \ , \qquad \mathcal{Y}_2 = \sqrt{\frac{\sqrt{1+\varkappa^2}-1}{\varkappa^2}} \ , \qquad \mathcal{Y}_3 = 0 \ , \qquad
\omega = \frac1{\sqrt{2}\mathcal{Y}_2} \ ,
\\
\label{sol5}
& \mathcal{Y}_1 = \sqrt[4]{1+\varkappa^2} \mathcal{Y}_3 \ , \qquad \mathcal{Y}_3 = \sqrt{\frac{\sqrt{1+\varkappa^2}-1}{\varkappa^2}} \ , \qquad \mathcal{Y}_2 = 0 \ , \qquad
\omega = \frac1{\sqrt{2}\mathcal{Y}_3} \ ,
\\
\label{sol6}
& \mathcal{Y}_2 = \sqrt[4]{1+\varkappa^2} \mathcal{Y}_3 \ , \qquad \mathcal{Y}_3 = \sqrt{\frac{\sqrt{1+\varkappa^2}-1}{\varkappa^2}} \ , \qquad \mathcal{Y}_1 = 0 \ , \qquad
\omega = \frac1{\sqrt{2}\mathcal{Y}_3} \ ,
\\ \nonumber
& \mathcal{Y}_1 = \sqrt{\frac{1 - 2\mathcal{Y}_3^2 - \varkappa^2 \mathcal{Y}_3^4}{1-\varkappa^2\mathcal{Y}_3^4}}
\ , \qquad
\mathcal{Y}_2 = \mathcal{Y}_3 \sqrt{\frac{1+\varkappa^2\mathcal{Y}_3^4}{1-\varkappa^2\mathcal{Y}_3^4}}
\ , \qquad
\omega = \sqrt{ \frac{1 + 2 \varkappa^2\mathcal{Y}_3^2 - \varkappa^2\mathcal{Y}_3^4} {1 + 3 \varkappa^2 \mathcal{Y}_3^4  - 2\varkappa^2\mathcal{Y}_3^6} }  \ ,
\\\label{sol7}
& \mathcal{Y}_3 = \sqrt{1 - \frac{\sqrt[3]{1+\varkappa^2}}{\varkappa} \big(e^{\frac{i\pi}{3}} \sqrt[3]{\varkappa+i} + e^{-\frac{i\pi}{3}} \sqrt[3]{\varkappa - i}\big) } \ ,
\end{align}
\endgroup
where we have omitted additional solutions related by signs to those presented here. We see that there are three special solutions, \eqref{sol1}, \eqref{sol2} and \eqref{sol3}, for which $\nu_{_S} = \omega = 1$ for all $\varkappa$. The characteristic feature of these solutions is the vanishing of the radii associated to two of the three angles $\varphi$, $\phi_1$ and $\phi_2$. Consequently the string is only moving in one of the Cartan directions.\foot{One should note that in these solutions the angles for which the corresponding radius is always vanishing could be set to any function of the worldsheet coordinates. Therefore, it does not make sense physically to give any such angles a non-trivial background. In particular, when considering the light-cone gauge action in section \ref{sec:smat} we set the background of these angles to zero.} It is these solutions for which we consider the associated light-cone gauge $S$-matrices in section \ref{sec:smat}.

Let us briefly remark on the remaining solutions. In the $\varkappa \to 0$ limit \eqref{sol4}, \eqref{sol5} and \eqref{sol6} reduce to
\begin{equation}\begin{split}
& \mathcal{Y}_1 = \mathcal{Y}_2 = \sqrt{\frac{1}{2}} \ , \qquad \mathcal{Y}_3 = 0 \ , \qquad \omega = 1 \ ,
\\ & \mathcal{Y}_1 = \mathcal{Y}_3 = \sqrt{\frac{1}{2}} \ , \qquad \mathcal{Y}_2 = 0 \ , \qquad \omega = 1 \ ,
\\ & \mathcal{Y}_2 = \mathcal{Y}_3 = \sqrt{\frac{1}{2}} \ , \qquad \mathcal{Y}_1 = 0 \ , \qquad \omega = 1 \ ,
\end{split}\end{equation}
while \eqref{sol7} becomes
\begin{equation}
\mathcal{Y}_{1} = \mathcal{Y}_2 = \mathcal{Y}_3 = \sqrt{\frac{1}{3}} \ , \qquad \omega = 1 \ .
\end{equation}
These solutions are defined for all $\varkappa \in \mathbb{R}$. However, care is needed when taking $\varkappa \to \pm \infty$ as in these limits $\omega \sim \sqrt{\varkappa}$. For a finite limit we first rescale the worldsheet time $\tau$ by $\frac1{\sqrt{\varkappa}}$, so that $\omega$ remains finite while $\nu_{_S} \to 0$.

\

Similar considerations can be made for the deformed $\AdS_5$ background. As described in section \ref{ssec:non-split} there are three possible deformations, which are related to \eqref{backemb} by the following analytic continuations
\begin{align}\label{rep00}
\mathcal{Y}_1  & \to \mathcal{Z}_0 \ ,
& \mathcal{Y}_{2,3} & \to i \mathcal{Z}_{1,2} \ ,
& \varphi & \to t \ ,
& \phi_1 & \to \psi_1 \ ,
& \phi_2 & \to \psi_2 \ ,
\\\label{rep11}
\mathcal{Y}_2 & \to \mathcal{Z}_0 \ ,
& \mathcal{Y}_{1,3} & \to i \mathcal{Z}_{1,2} \ ,
& \varphi & \to \psi_1 \ ,
& \phi_1 & \to t \ ,
& \phi_2 & \to \psi_2 \ ,
\\\label{rep22}
\mathcal{Y}_3 & \to \mathcal{Z}_0 \ ,
& \mathcal{Y}_{1,2} & \to i \mathcal{Z}_{1,2} \ ,
& \varphi & \to \psi_2 \ ,
& \phi_1 & \to \psi_1 \ ,
& \phi_2 & \to t \ ,
\end{align}
along with reversing the overall sign of the metric and $B$-field.\foot{\label{footmapa}After
using the replacement \eqref{rep00} in \eqref{backemb}, to recover \eqref{backk0} up to signs one should set
$\mathcal{Z}_0 = \sqrt{1+\rho^2}$, $\mathcal{Z}_1 = \rho \sqrt{1-x^2}$ and $\mathcal{Z}_2 = \rho x$.
Similarly, after using the replacements \eqref{rep11} or \eqref{rep22} in \eqref{backemb}, we set
$\mathcal{Z}_0 = \sqrt{1+\rho^2}\sqrt{1+x^2}$, $\mathcal{Z}_1 = \rho$ and $\mathcal{Z}_2 = \sqrt{1+\rho^2} x$
to recover \eqref{backk1} or \eqref{backk2} respectively.} In the $\varkappa \to 0$ limit
these coordinates are again easily related to the embedding coordinates \eqref{embeddc}
\begin{equation}
Z_0 + i Z_5 = \mathcal{Z}_0 \, e^{i t} \ , \qquad
Z_1 + i Z_2 = \mathcal{Z}_1 \, e^{i \psi_1} \ , \qquad
Z_3 + i Z_4 = \mathcal{Z}_2 \, e^{i \psi_2} \ . \qquad
\end{equation}

Following the same derivation as for the deformed $S^5$, we find that for each of the three possible deformations there is only a single solution of interest
\begin{equation}\label{lista}
\mathcal{Z}_0 = 1 \ , \qquad \mathcal{Z}_{1,2} = 0 \ , \qquad t = \tau \ .
\end{equation}
These are the solutions for which we consider the associated light-cone gauge $S$-matrices in section \ref{sec:smat}.
The contribution from the deformed $\AdS$ part of the geometry to the conformal-gauge Virasoro constraints is
\begin{equation}\label{vira}
\nu_{_A} = G^A_{\mu\nu} \partial_\pm X^\mu \partial_\pm X^\nu = - 1 \ ,
\end{equation}
where $G^A_{\mu\nu}$ is the metric of the deformed $\AdS_5$. When classifying solutions in \eqref{lista} we have imposed the additional requirement that $\nu_{_A} \leq 0$. Therefore, by rescaling the dependence on the worldsheet time in the classical solution, we can arrange that $\nu_{_A} = -1$. This means that we can pair any the solutions on the deformed $\AdS_5$ \eqref{lista} with any of those on the deformed $S^5$ \eqref{sol1} -- \eqref{sol7} to solve the conformal-gauge Virasoro constraints
\begin{equation}
0 = G_{\mu\nu}\partial_\pm X^\mu \partial_\pm X^\nu = \nu_{_A} + \nu_{_S} \ .
\end{equation}
This classifies the string solutions of interest.



\end{document}